\pdfoutput=1

\documentclass[12pt,a4paper]{article}

\usepackage{ifthen} 
\newboolean{pdflatex}
\setboolean{pdflatex}{true} 

\newboolean{articletitles}
\setboolean{articletitles}{true} 

\newboolean{uprightparticles}
\setboolean{uprightparticles}{true} 

\newboolean{inbibliography}
\setboolean{inbibliography}{false} 

\usepackage{rotating}  %
\usepackage{graphpap}  %
\usepackage{multirow}  %
\usepackage{mathrsfs}  
\usepackage{pifont}    
\usepackage{afterpage} 

\newcommand*\diff{\mathop{}\!\mathrm{d}}

\usepackage{microtype}
\usepackage{lineno}  
\usepackage{xspace} 
\usepackage{caption} 

\usepackage{graphicx}  
\usepackage{color}
\usepackage{colortbl}
\graphicspath{{./figs/}} 

\usepackage{amsmath} 
\usepackage{amssymb}
\usepackage{amsfonts}
\usepackage{upgreek} 

\newcommand*\patchAmsMathEnvironmentForLineno[1]{%
\expandafter\let\csname old#1\expandafter\endcsname\csname #1\endcsname
\expandafter\let\csname oldend#1\expandafter\endcsname\csname
end#1\endcsname
 \renewenvironment{#1}%
   {\linenomath\csname old#1\endcsname}%
   {\csname oldend#1\endcsname\endlinenomath}%
}
\newcommand*\patchBothAmsMathEnvironmentsForLineno[1]{%
  \patchAmsMathEnvironmentForLineno{#1}%
  \patchAmsMathEnvironmentForLineno{#1*}%
}
\AtBeginDocument{%
\patchBothAmsMathEnvironmentsForLineno{equation}%
\patchBothAmsMathEnvironmentsForLineno{align}%
\patchBothAmsMathEnvironmentsForLineno{flalign}%
\patchBothAmsMathEnvironmentsForLineno{alignat}%
\patchBothAmsMathEnvironmentsForLineno{gather}%
\patchBothAmsMathEnvironmentsForLineno{multline}%
\patchBothAmsMathEnvironmentsForLineno{eqnarray}%
}

\usepackage{hyperref}    
\usepackage[all]{hypcap} 

\def\lhcb {\mbox{LHCb}\xspace}

\def\rich   {RICH\xspace}

\def\MagUp {\mbox{\em Mag\kern -0.05em Up}\xspace}

\ifthenelse{\boolean{uprightparticles}}%
{

 \def\Peta        {\ensuremath{\upeta}\xspace}

 \def\Pmu         {\ensuremath{\upmu}\xspace}                 
 \def\Pnu         {\ensuremath{\upnu}\xspace}                 
                  
 \def\Ppi         {\ensuremath{\uppi}\xspace}

 \def\Ppsi        {\ensuremath{\uppsi}\xspace}

 \def\PDelta      {\ensuremath{\Delta}\xspace}                 
 \def\PXi      {\ensuremath{\Xi}\xspace}                 
 \def\PLambda      {\ensuremath{\Lambda}\xspace}                 
 \def\PSigma      {\ensuremath{\Sigma}\xspace}                 
 \def\POmega      {\ensuremath{\Omega}\xspace}                 
 \def\PUpsilon      {\ensuremath{\Upsilon}\xspace}                 
 
 \def\PB      {\ensuremath{\mathrm{B}}\xspace}                 
                  
 \def\PD      {\ensuremath{\mathrm{D}}\xspace}

 \def\PJ      {\ensuremath{\mathrm{J}}\xspace}                 
 \def\PK      {\ensuremath{\mathrm{K}}\xspace}

 \def\Pb      {\ensuremath{\mathrm{b}}\xspace}                 
 \def\Pc      {\ensuremath{\mathrm{c}}\xspace}

 \def\Pi      {\ensuremath{\mathrm{i}}\xspace}

 \def\Pp      {\ensuremath{\mathrm{p}}\xspace}

 \def\Ps      {\ensuremath{\mathrm{s}}\xspace}

}
{

 \def\Peta        {\ensuremath{\eta}\xspace}

 \def\Pmu         {\ensuremath{\mu}\xspace}                 
 \def\Pnu         {\ensuremath{\nu}\xspace}                 
                  
 \def\Ppi         {\ensuremath{\pi}\xspace}

 \def\Ppsi        {\ensuremath{\psi}\xspace}                 
                  
 \mathchardef\PDelta="7101
 \mathchardef\PXi="7104
 \mathchardef\PLambda="7103
 \mathchardef\PSigma="7106
 \mathchardef\POmega="710A
 \mathchardef\PUpsilon="7107
                  
 \def\PB      {\ensuremath{B}\xspace}                 
                  
 \def\PD      {\ensuremath{D}\xspace}

 \def\PJ      {\ensuremath{J}\xspace}                 
 \def\PK      {\ensuremath{K}\xspace}

 \def\Pb      {\ensuremath{b}\xspace}                 
 \def\Pc      {\ensuremath{c}\xspace}

 \def\Pi      {\ensuremath{i}\xspace}

 \def\Pp      {\ensuremath{p}\xspace}

 \def\Ps      {\ensuremath{s}\xspace}

}

\makeatletter
\ifcase \@ptsize \relax
  \newcommand{\miniscule}{\@setfontsize\miniscule{4}{5}}
\or
  \newcommand{\miniscule}{\@setfontsize\miniscule{5}{6}}
\or
  \newcommand{\miniscule}{\@setfontsize\miniscule{5}{6}}
\fi
\makeatother

\DeclareRobustCommand{\optbar}[1]{\shortstack{{\miniscule (\rule[.5ex]{1.25em}{.18mm})}
  \\ [-.7ex] $#1$}}

\def\mup        {{\ensuremath{\Pmu^+}}\xspace}

\def\squark    {{\ensuremath{\Ps}}\xspace}

\def\cquark    {{\ensuremath{\Pc}}\xspace}

\def\bquark    {{\ensuremath{\Pb}}\xspace}

\def\pion   {{\ensuremath{\Ppi}}\xspace}

\def\pip    {{\ensuremath{\pion^+}}\xspace}

\def\kaon    {{\ensuremath{\PK}}\xspace}
  \def\Kbar    {{\kern 0.2em\overline{\kern -0.2em \PK}{}}\xspace}

\def\KorKbar    {\kern 0.18em\optbar{\kern -0.18em K}{}\xspace}

\def\Kp      {{\ensuremath{\kaon^+}}\xspace}

  \def\Dbar    {{\kern 0.2em\overline{\kern -0.2em \PD}{}}\xspace}

\def\DorDbar    {\kern 0.18em\optbar{\kern -0.18em D}{}\xspace}

\def\B       {{\ensuremath{\PB}}\xspace}
\def\Bbar    {{\ensuremath{\kern 0.18em\overline{\kern -0.18em \PB}{}}}\xspace}

\def\BorBbar    {\kern 0.18em\optbar{\kern -0.18em B}{}\xspace}

\def\Bu      {{\ensuremath{\B^+}}\xspace}

\def\Bd      {{\ensuremath{\B^0}}\xspace}
\def\Bs      {{\ensuremath{\B^0_\squark}}\xspace}

\def\Bc      {{\ensuremath{\B_\cquark^+}}\xspace}

\def\jpsi     {{\ensuremath{{\PJ\mskip -3mu/\mskip -2mu\Ppsi\mskip 2mu}}}\xspace}

  \def\Y#1S{\ensuremath{\PUpsilon{(#1S)}}\xspace}

\def\proton      {{\ensuremath{\Pp}}\xspace}

\def\Lbar        {{\ensuremath{\kern 0.1em\overline{\kern -0.1em\PLambda}}}\xspace}
\def\LorLbar    {\kern 0.18em\optbar{\kern -0.18em \PLambda}{}\xspace}

\def\to                 {\ensuremath{\rightarrow}\xspace}

\newcommand{\lqcd}{{\ensuremath{\Lambda_{\mathrm{QCD}}}}\xspace}

\def\AT#1     {\ensuremath{A_{\mathrm{T}}^{#1}}\xspace}           

\def\C#1      {\ensuremath{\mathcal{C}_{#1}}\xspace}                       
\def\Cp#1     {\ensuremath{\mathcal{C}_{#1}^{'}}\xspace}                    
\def\Ceff#1   {\ensuremath{\mathcal{C}_{#1}^{\mathrm{(eff)}}}\xspace}        
\def\Cpeff#1  {\ensuremath{\mathcal{C}_{#1}^{'\mathrm{(eff)}}}\xspace}       
\def\Ope#1    {\ensuremath{\mathcal{O}_{#1}}\xspace}                       
\def\Opep#1   {\ensuremath{\mathcal{O}_{#1}^{'}}\xspace}                    

\newcommand{\tev}{\ifthenelse{\boolean{inbibliography}}{\ensuremath{~T\kern -0.05em eV}\xspace}{\ensuremath{\mathrm{\,Te\kern -0.1em V}}}\xspace}
\newcommand{\gev}{\ensuremath{\mathrm{\,Ge\kern -0.1em V}}\xspace}
\newcommand{\mev}{\ensuremath{\mathrm{\,Me\kern -0.1em V}}\xspace}
\newcommand{\kev}{\ensuremath{\mathrm{\,ke\kern -0.1em V}}\xspace}
\newcommand{\ev}{\ensuremath{\mathrm{\,e\kern -0.1em V}}\xspace}
\newcommand{\gevc}{\ensuremath{{\mathrm{\,Ge\kern -0.1em V\!/}c}}\xspace}
\newcommand{\mevc}{\ensuremath{{\mathrm{\,Me\kern -0.1em V\!/}c}}\xspace}
\newcommand{\gevcc}{\ensuremath{{\mathrm{\,Ge\kern -0.1em V\!/}c^2}}\xspace}
\newcommand{\gevgevcccc}{\ensuremath{{\mathrm{\,Ge\kern -0.1em V^2\!/}c^4}}\xspace}
\newcommand{\mevcc}{\ensuremath{{\mathrm{\,Me\kern -0.1em V\!/}c^2}}\xspace}

\def\mm   {\ensuremath{\rm \,mm}\xspace}

\def\mum  {\ensuremath{{\,\upmu\rm m}}\xspace}

\def\invfb   {\ensuremath{\mbox{\,fb}^{-1}}\xspace}

\def\ps   {\ensuremath{{\rm \,ps}}\xspace}
\def\fs   {\ensuremath{\rm \,fs}\xspace}

\newcommand{\chisq}{\ensuremath{\chi^2}\xspace}

\def\gsim{{~\raise.15em\hbox{$>$}\kern-.85em
          \lower.35em\hbox{$\sim$}~}\xspace}
\def\lsim{{~\raise.15em\hbox{$<$}\kern-.85em
          \lower.35em\hbox{$\sim$}~}\xspace}

\def\ptot       {\mbox{$p$}\xspace}
\def\pt         {\mbox{$p_{\rm T}$}\xspace}

\def\bcvegpy    {\mbox{\textsc{Bcvegpy}}\xspace}

\def\evtgen     {\mbox{\textsc{EvtGen}}\xspace}

\def\geant      {\mbox{\textsc{Geant4}}\xspace}

\def\photos     {\mbox{\textsc{Photos}}\xspace}

\def\pythia     {\mbox{\textsc{Pythia}}\xspace}

\def\tell1  {TELL1\xspace}
\def\ukl1   {UKL1\xspace}

\newcommand{\ie}{\mbox{\itshape i.e.}\xspace}

\usepackage{cite} 
\usepackage{mciteplus}

\usepackage{longtable} 

\begin{document}

\renewcommand{\thefootnote}{\fnsymbol{footnote}}
\setcounter{footnote}{1}


\begin{titlepage}
\pagenumbering{roman}

\vspace*{-1.5cm}
\centerline{\large EUROPEAN ORGANIZATION FOR NUCLEAR RESEARCH (CERN)}
\vspace*{1.5cm}
\hspace*{-0.5cm}
\begin{tabular*}{\linewidth}{lc@{\extracolsep{\fill}}r}
\ifthenelse{\boolean{pdflatex}}
{\vspace*{-2.7cm}\mbox{\!\!\!\includegraphics[width=.14\textwidth]{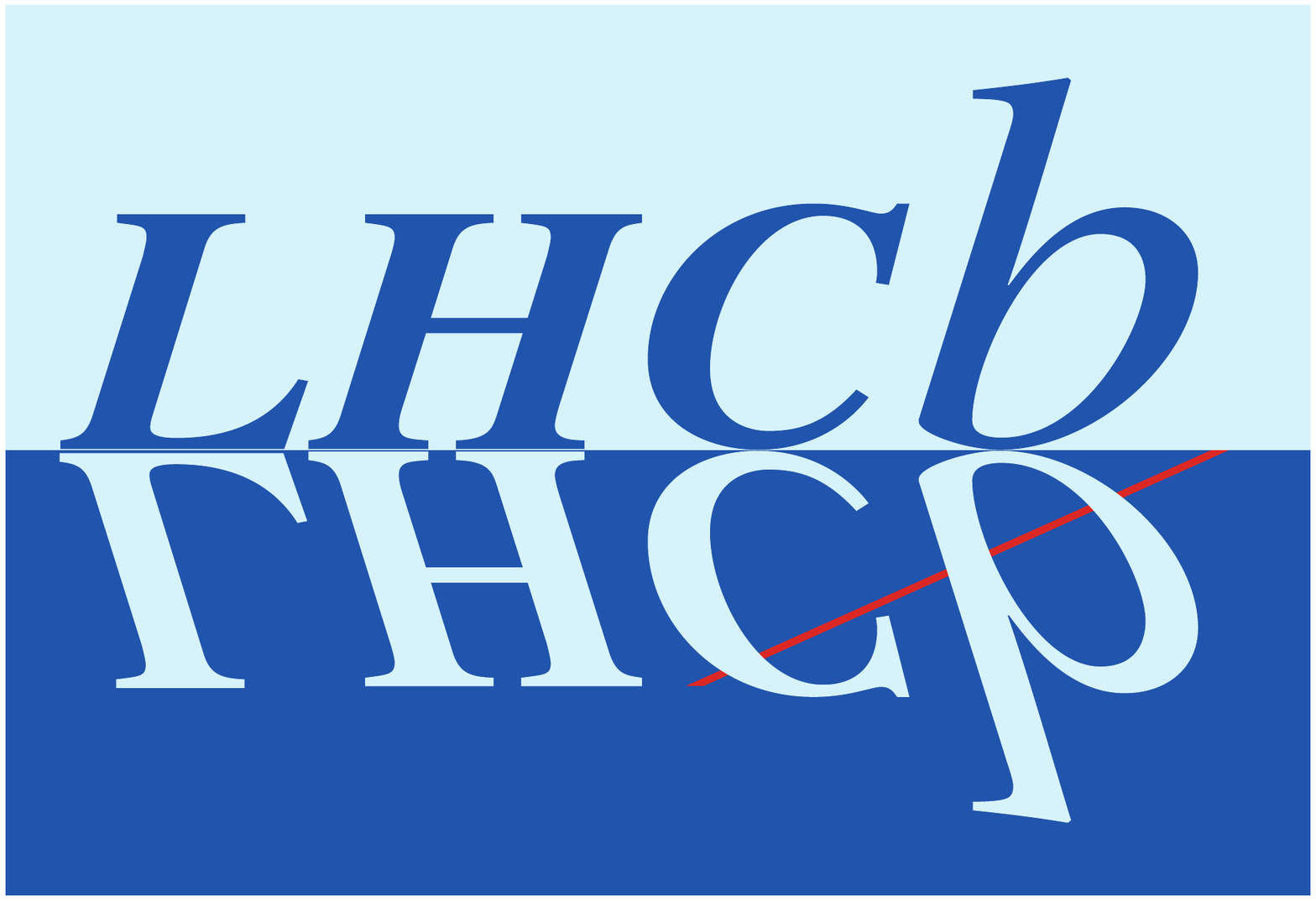}} & &}%
{\vspace*{-1.2cm}\mbox{\!\!\!\includegraphics[width=.12\textwidth]{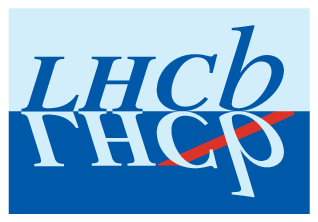}} & &}%
\\
 & & CERN-PH-EP-2014-283 \\  
 & & LHCb-PAPER-2014-060 \\  
 & & November 25, 2014 \\ 
 & & \\
\end{tabular*}

\vspace*{1.0cm}

{\bf\boldmath\huge
  \begin{center}
    Measurement of the~lifetime of the~\Bc~meson using
    the~$\Bc\to\jpsi\pip$~decay mode
  \end{center}
}

\vspace*{0.7cm}

\begin{center}
The LHCb collaboration\footnote{Authors are listed at the end of this letter.}
\end{center}

\vspace{\fill}

\begin{abstract}
  \noindent
  The~difference in total widths between the~\Bc~and \Bu~mesons  is measured 
  using 3.0\invfb of data collected 
  by the LHCb experiment  in 7~and 8\tev centre-of-mass 
  energy proton-proton~collisions at the~LHC.
  Through the~study of the~time evolution of 
  \mbox{$\Bc\to\jpsi\pip$}~and 
  \mbox{$\Bu\to\jpsi\Kp$}~decays, 
  the~width difference is measured to be 
  \begin{equation*}
    \Delta\Gamma \equiv \Gamma_{\Bc} - \Gamma_{\Bu} =  
      4.46 \pm 0.14  \pm0.07 \mm^{-1}  {\mathit{c}}, 
  \end{equation*}
  where the first uncertainty is statistical and 
  the~second systematic. 
  The~known lifetime of the~\Bu~meson is
  used to convert this to a precise measurement of the~\Bc~lifetime,
  \begin{equation*}
   \uptau_{\Bc}    =    513.4 \pm 11.0           \pm 5.7\fs,   
  \end{equation*}
  where the~first uncertainty is statistical 
  and the~second systematic.
\end{abstract}

\vspace*{0.7cm}

\begin{center}
  Submitted to Phys.~Lett.~B
\end{center}

\vspace{\fill}

{\footnotesize 
  \centerline{\copyright~CERN on behalf of the~\lhcb collaboration,
    licence \href{http://creativecommons.org/licenses/by/4.0/}{CC-BY-4.0}.}}
\vspace*{2mm}

\end{titlepage}


\newpage
\setcounter{page}{2}
\mbox{~}
%
%
%
%

\cleardoublepage


\renewcommand{\thefootnote}{\arabic{footnote}}
\setcounter{footnote}{0}



\pagestyle{plain} 
\setcounter{page}{1}
\pagenumbering{arabic}


%

\section{Introduction}
\label{sec:intro}
For weakly decaying beauty hadrons
the heavy quark expansion~\cite{Khoze:1983yp,Bigi:1991ir,Blok:1992hw,*Blok:1992he,Lenz:2014jha}
predicts lifetime differences of the~order $\left(\lqcd/m_{\bquark}\right)$, 
where \lqcd~is the~scale parameter of the strong interaction  
and $m_{\bquark}$ is the \bquark-quark mass.
In agreement with the~expectations, differences
between
\Bu, \Bd, \Bs,
$\PLambda_{\bquark}^{0}$,
$\PXi_{\bquark}^{0}$~and
$\PXi_{\bquark}^{-}$~lifetimes 
not exceeding
a~few per cent are found experimentally~\cite{PDG2014,
  LHCb-PAPER-2013-065,
  LHCb-PAPER-2013-032,
  LHCb-PAPER-2014-003,
  LHCb-PAPER-2014-021,
  LHCb-PAPER-2014-037,
  LHCb-PAPER-2014-048}.
The~\Bc~meson
is a~bound state of an anti-\bquark~quark and a~charm~quark, 
and Cabibbo-favoured decays of the~charm~quark are
expected to account for 70\,\% of its~total width,
resulting in a~significantly shorter lifetime than for~other B mesons.
In~addition, non-spectator topologies, 
in particular annihilation amplitudes,
are not suppressed. 
These could give 
sizeable contributions 
to the~total width~\cite{Chang:1992pt,Gershtein:1994jw,Colangelo:1999zn,Kiselev:2000pp,*Kiselev:2001ej,Chang:2000ac,*Chang:2001jn,Kiselev:2003mp}.
Understanding the~relative contributions of beauty and charm quarks to
the~total width of the~\Bc~meson is important for predicting
the~properties of unobserved baryons with two heavy quarks~\cite{Kiselev:2001fw,Karliner:2014gca}.

The~lifetime of the~\Bc~meson was first measured by 
the~CDF~\cite{Abe:1998wi,Abulencia:2006zu,Aaltonen:2012yb}
and D0~\cite{Abazov:2008rba} collaborations
using semileptonic $\Bc\to\jpsi\mup\Pnu_{\upmu}\mathrm{X}$
and hadronic $\Bc\to\jpsi\pip$~decays.
The~average value of these measurements is
\mbox{$\uptau_{\Bc}=452\pm32\fs$}~\cite{PDG2012}.
Recently, the~LHCb collaboration made the~most precise measurement
to date of the \Bc~meson lifetime using semileptonic $\Bc\to\jpsi\mup\Pnu_{\upmu}\mathrm{X}$~decays, 
\mbox{$\uptau_{\Bc} = 509\pm14\fs$}~\cite{LHCb-PAPER-2013-063}. 

In this Letter we report a measurement
of the \Bc~meson lifetime obtained via
the~difference between the total width 
of the~\Bc~and \Bu~mesons in the~hadronic modes $\Bc\to\jpsi\pip$
and $\Bu\to\jpsi\Kp$, using the~technique
developed in Refs.~\cite{LHCb-PAPER-2012-017,
  LHCb-PAPER-2013-032,
  LHCb-PAPER-2014-003,
  LHCb-PAPER-2014-021,
  LHCb-PAPER-2014-037,
  LHCb-PAPER-2014-048}.  The measurement uses 3.0\invfb of data collected 
by the~LHCb experiment in proton-proton\,($\proton\proton$)~collisions
at centre-of-mass energies of 7~and 8\tev. This study is complementary
to the~measurement 
of the~\Bc~lifetime using the~semileptonic 
$\Bc\to\jpsi\mup\Pnu_{\upmu}\mathrm{X}$~decays described
in Ref.~\cite{LHCb-PAPER-2013-063}.

The \Bc~lifetime is determined as follows.
The decay time distribution for signal, $N_{\B}\left(t\right)$, can be 
described as the product of an acceptance function $\varepsilon_{\B}\left(t\right)$ 
and an~exponential decay $E_{\B}\left(t\right) = \exp\left(-t/\uptau_{\B}\right)$
convolved 
with the~decay time resolution of the~detector.
The~effect of the~decay time resolution on the~ratio
$\mathcal{R}\left(t\right) \equiv N_{\Bc}\left(t\right)/N_{\Bu}\left(t\right)$
is found to be small and is absorbed into
the~ratio of acceptance functions.
This leads to the simplified expression 
\begin{align}
\begin{split}
  \mathcal{R}\left(t\right)
  & \propto   
  \dfrac{\varepsilon_{\Bc}\left(t\right)}{\varepsilon_{\Bu}\left(t\right)}
  \dfrac{E_{\uptau_{\Bc}}\left(t\right)}{E_{\uptau_{\Bu}}\left(t\right)}
  \equiv 
  \mathcal{R}_{\varepsilon}\left(t\right) \, e^{-\Delta\Gamma t}  \\ 
 \text{with}~~~~
 \Delta\Gamma  & \equiv  \Gamma_{\Bc} - \Gamma_{\Bu}
 =  \dfrac{1}{\uptau_{\Bc}} - \dfrac{1}{\uptau_{\B}}\;,  \label{eq:rd} 
\end{split}
\end{align}
where the factor $R_\varepsilon\left(t\right)$
denotes the~ratio of the~acceptance functions.
This allows a~precise measurement of 
$\Delta\Gamma$~and hence of the lifetime of the~\Bc~meson.

\section{Detector and event simulation}
\label{sec:Detector}

The \lhcb detector~\cite{Alves:2008zz} is a single-arm forward
spectrometer covering the \mbox{pseudorapidity} range $2<\Peta <5$,
designed for the study of particles containing \bquark or \cquark
quarks. The~detector includes a~high-precision tracking system
consisting of a silicon-strip vertex detector surrounding 
the $\proton\proton$~interaction region~\cite{LHCb-DP-2014-001}, 
a large-area silicon-strip detector located
upstream of a~dipole magnet with a bending power of about
$4{\rm\,Tm}$, and three stations of silicon-strip detectors and straw
drift tubes~\cite{LHCb-DP-2013-003} placed downstream of the~magnet.
The~tracking system provides a~measurement of momentum, \ptot,  with
a~relative uncertainty that varies 
from 0.4\% at low momentum to 0.6\% at 100\gevc.
The minimum distance of a~track to a~primary vertex, 
the~impact parameter, is measured with a~resolution of $(15+29/\pt)\mum$,
where \pt is the~component of momentum transverse to the~beam, in \gevc.
Different types of charged hadrons are distinguished using information
from two ring-imaging Cherenkov detectors\,(\rich)~\cite{LHCb-DP-2012-003}. 
Photon, electron and
hadron candidates are identified by a~calorimeter system consisting of
scintillating-pad and preshower detectors, an~electromagnetic
calorimeter and a~hadronic calorimeter. Muons are identified
by a~system composed of alternating layers of iron and multiwire
proportional chambers~\cite{LHCb-DP-2012-002}.

The trigger~\cite{LHCb-DP-2012-004} comprises two stages.
Events are first required to pass the~hardware trigger, 
which selects
muon candidates with $\pt>1.5\gevc$ or
pairs of opposite-sign muon candidates 
with a~requirement that the~product of the~muon transverse momenta 
is larger than $1.7\,(2.6)\,\mathrm{GeV}^2/c^2$ 
for data collected at $\sqrt{s}=7\,(8)\tev$.
The subsequent software trigger is composed of two stages, 
the~first of which performs a~partial event reconstruction,
while full event reconstruction is done at the~second stage.
%
At the~first stage of the~software trigger 
the invariant mass of well-reconstructed pairs of 
oppositely charged muons forming a~good two-prong vertex 
is required to exceed 2.7\gevcc,
and the~two-prong vertex 
is required to be significantly displaced 
with respect to the reconstructed \proton\proton~collision vertex. 

In the simulation, $\proton\proton$ collisions 
are generated using \pythia~\cite{Sjostrand:2006za,*Sjostrand:2007gs} 
with a specific \lhcb~configuration~\cite{LHCb-PROC-2010-056}.  
A~dedicated generator,
\bcvegpy~\cite{Chang:2003cq,*Chang:2005hq,*Wu:2013pya}, 
which implements explicit 
leading-order matrix element
calculations~\cite{Berezhnoy:1994ba,Kolodziej:1995nv,Chang:1994aw},
is used for production of \Bc~mesons.
The~kinematic distributions of \Bc~mesons 
are reproduced by
the~\bcvegpy~generator with percent-level 
precision~\cite{LHCb-PAPER-2013-010,
  LHCb-PAPER-2013-021,
  LHCb-PAPER-2013-044,
  LHCb-PAPER-2013-047,
  LHCb-PAPER-2013-063,
  LHCb-PAPER-2014-009,
  LHCb-PAPER-2014-025,
  LHCb-PAPER-2014-039,
  LHCb-PAPER-2014-050},
while
the~simulated \Bu~samples, produced with \pythia, 
are corrected to reproduce the~observed kinematic 
distributions.
Decays of hadronic particles are described by \evtgen~\cite{Lange:2001uf}, 
in which final-state radiation is generated using \photos~\cite{Golonka:2005pn}. 
The~interaction of the generated particles with 
the~detector and the~detector response
are implemented using
the~\geant
toolkit~\cite{Allison:2006ve, *Agostinelli:2002hh} as described in
Ref.~\cite{LHCb-PROC-2011-006}.


\section{Event Selection}
\label{sec:evsel}
The offline selection of $\Bc\to\jpsi\pip$ and
$\Bu\to\jpsi\Kp$~candidates is divided into two parts.
An~initial  selection is applied to reduce the combinatorial background.
Subsequently, a~multivariate estimator based 
on an~artificial neural network algorithm~\cite{McCulloch,rosenblatt58}, 
configured with a~cross-entropy cost estimator~\cite{Zhong:2011xm}, 
in the following referred as $\mathtt{MLP}$~classifier, is applied.
The same criteria are used for both the  \Bc~and \Bu~candidates.
 
The selection starts from well-identified muon candidates that have
a~transverse momentum in~excess of 550\mevc.
Pairs of muon candidates are required
to form a~common vertex and
to have an~invariant mass within $\pm60\mevcc$ of 
the~known \jpsi~mass~\cite{PDG2014}.
To~ensure that the~\jpsi~candidate
originates in a~\bquark-hadron decay,
a~significant decay length with
respect to the \proton\proton~collision vertex is required. 
The charged pions and kaons must be positively identified using the combined
information from the~\rich, calorimeter and muon detectors.
The~\Bc~and  \Bu~candidates are formed from $\jpsi\pip$~and
$\jpsi\Kp$~combinations, respectively. 
To~improve the invariant mass and decay~time resolutions for selected candidates, 
a~kinematic fit~\cite{Hulsbergen:2005pu} is applied
in which
a~primary vertex pointing constraint and a~mass constraint on
the~intermediate \jpsi~states are applied.
To~reduce combinatorial background,
a~requirement on
the~$\chi^2$ of this fit, $\chi^2_{\mathrm{fit}}$, is imposed and
 the~decay time of the~reconstructed $\Bc\,(\Bu)$~candidate
 is required to be in the~range $50< t <1000\mum/c$.

The final selection of candidates using the $\mathtt{MLP}$~classifier
is based on the transverse momenta and rapidities of 
reconstructed $\Bc\,(\Bu)$ and \jpsi, 
the transverse momentum and pseudorapidity of 
the~$\pip\,(\Kp)$~candidate, 
the~cosine of the~decay angle $\theta$ between  
the~momentum of 
the~$\pip\,(\Kp)$ in the~rest frame of
the~$\Bc\,(\Bu)$~candidate 
and the boost direction from the $\Bc\,(\Bu)$~rest frame 
to the~laboratory frame, the $\chisq$ of 
the~$\Bc\,(\Bu)$~vertex fit, and $\chi^2_{\mathrm{fit}}$.
These variables provide good discrimination between
signal and background whilst keeping the~selection efficiency 
independent of the~$\Bc\,(\Bu)$~decay time.
The~$\mathtt{MLP}$~classifier is trained on a simulated sample 
of $\Bc\to\jpsi\pip$~events and a background data sample 
from the mass sidebands of the~\Bc~signal peak.
It is tested on independent samples from the same sources. 
The~working point of the~classifier is chosen to minimize
$\upsigma({\mathcal{S}})/{\mathcal{S}}$, 
where $\mathcal{S}$ is the \Bc~signal yield 
and $\upsigma({\mathcal{S}})$ is 
the~yield  uncertainty, as determined
by the~mass fit described in the~next section.
The~same $\mathtt{MLP}$~classifier is used for 
the \mbox{$\Bu\to\jpsi\Kp$}~mode.

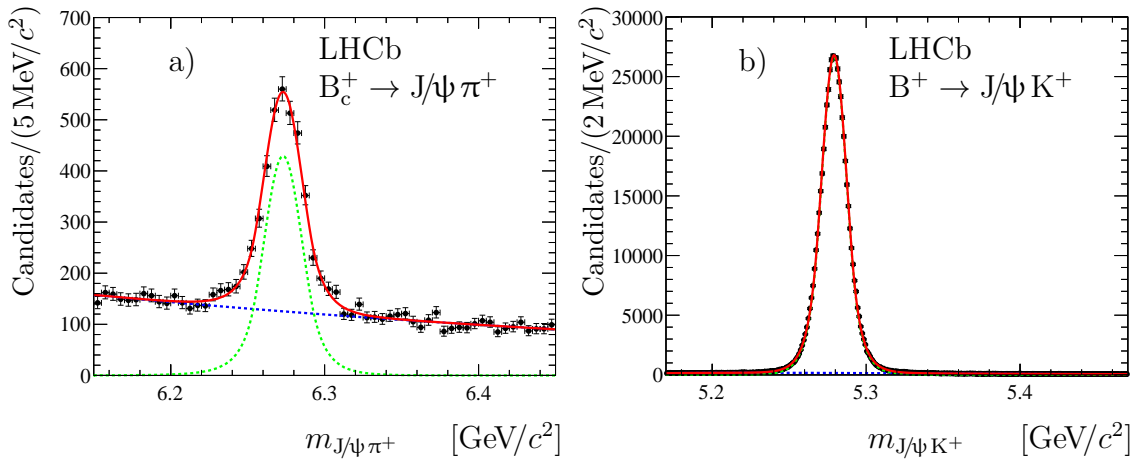
\begin{figure}[t!]
  \setlength{\unitlength}{1mm}
  \centering
  \begin{picture}(150,60)
    \put( 0, 0){
      \includegraphics*[width=75mm,height=60mm,%
      ]{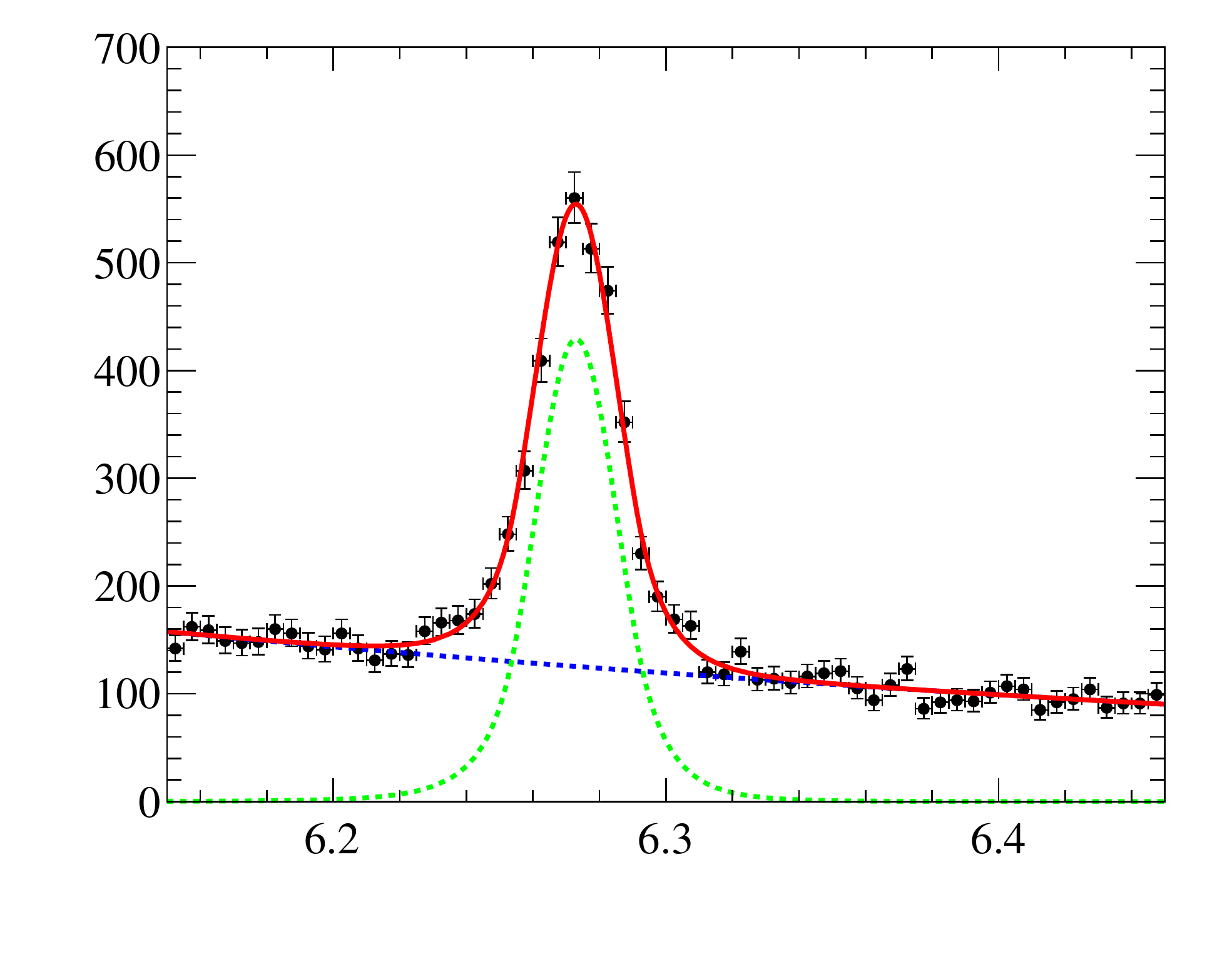}
    }
    \put(75, 0){
      \includegraphics*[width=75mm,height=60mm,%
      ]{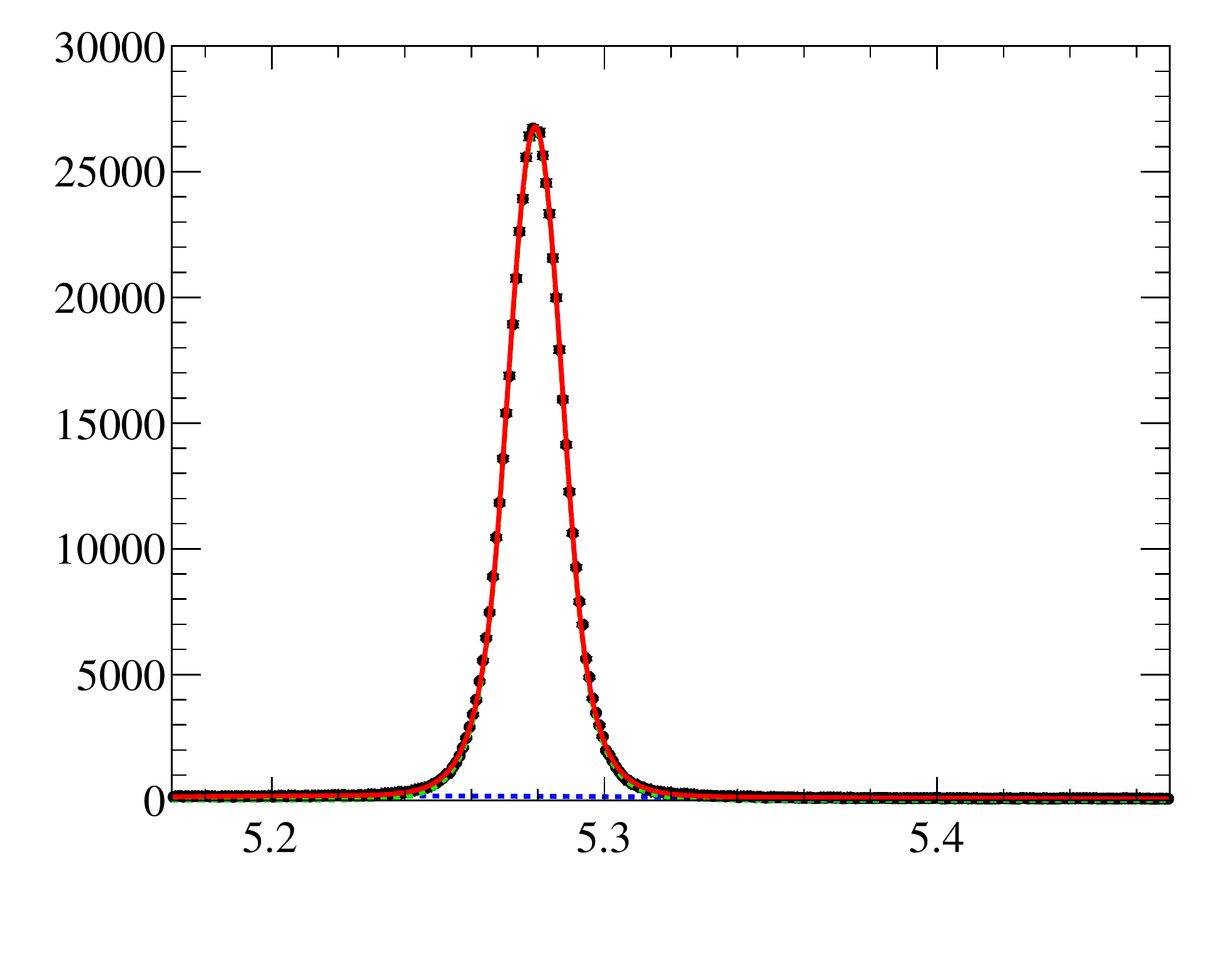}
    }
    \put( 20, 50){ a)}
    \put( 95, 50){ b)}
    \put( 38, 49){ $\begin{array}{l}\mathrm{LHCb} \\  \Bc\to\jpsi\pip \end{array}$ }
    \put(113, 49){ $\begin{array}{l}\mathrm{LHCb} \\  \Bu\to\jpsi\Kp   \end{array}$ }
    \put( -1, 20) { \small \begin{sideways}Candidates/(5\mevcc)\end{sideways} }
    \put( 74, 20) { \small \begin{sideways}Candidates/(2\mevcc)\end{sideways} }
    \put( 38,  0) { $m_{\jpsi\pip}$ }
    \put(112,  0) { $m_{\jpsi\Kp}$  }
    \put( 57,  0) { $\left[\!\gevcc\right]$ }
    \put(132,  0) { $\left[\!\gevcc\right]$ }
  \end{picture}
  \caption { \small
    Invariant mass distributions for (a)~selected 
    $\Bc\to\jpsi\pip$  and
    (b)~$\Bu\to\jpsi\Kp$~candidates.
    The fit result with the~function described
    in the~text is shown 
    by the~red solid line; the~signal\,(background) components 
    are shown with green\,(blue)  dotted\,(dashed) lines.
  }
  \label{fig:signals}
\end{figure}

{\boldmath{\section{Measurement of $\Delta\Gamma$}}}
\label{sec:time}
The invariant mass distributions for 
selected \Bc and \Bu~candidates are presented 
in Fig.~\ref{fig:signals}. 
The~signal yields are determined using an~extended unbinned maximum likelihood fit
in which the~signal distributions are modelled by 
a~Gaussian function with power-law 
tails on both sides of the peak~\cite{LHCb-PAPER-2011-013}, 
and the background is modelled by the product
of an~exponential  and a~first-order polynomial function.
Simulation studies suggest that
the~same tail parameters apply for the \Bc and \Bu~signals.
The tail parameters determined from the data for the large \Bu~signal are
in good agreement with the simulation.
The fit gives 
$2886     \pm 71$~signal \Bc~decays and 
$586\,065 \pm 798$~signal \Bu~decays.
The fitted values for the \Bc and \Bu~invariant masses are consistent with the known values~\cite{PDG2014} and 
the fitted mass~resolutions agree with the~expectation from simulation.

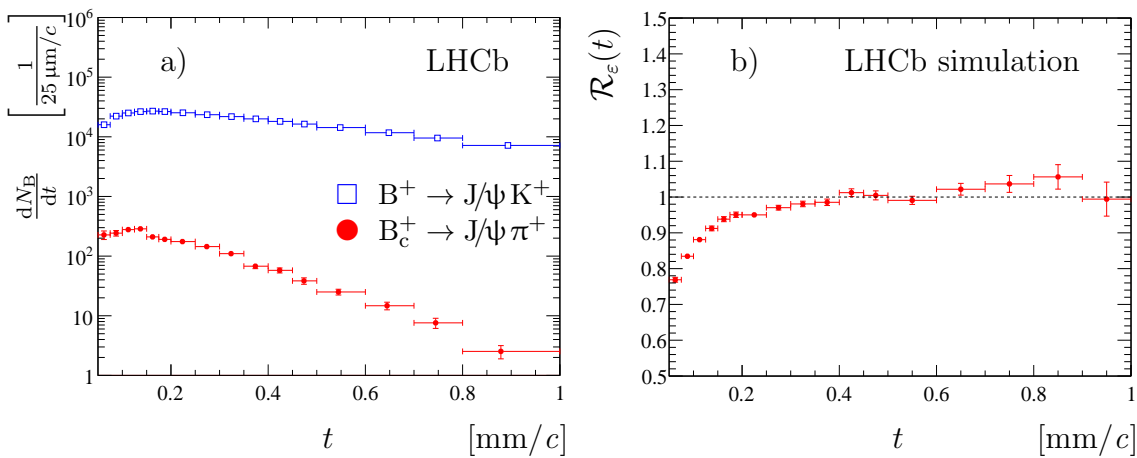
\begin{figure}[t!]
  \setlength{\unitlength}{1mm}
  \centering
  \begin{picture}(150,60)
    \put(  0,  0){
      \includegraphics*[width=75mm,height=60mm,%
      ]{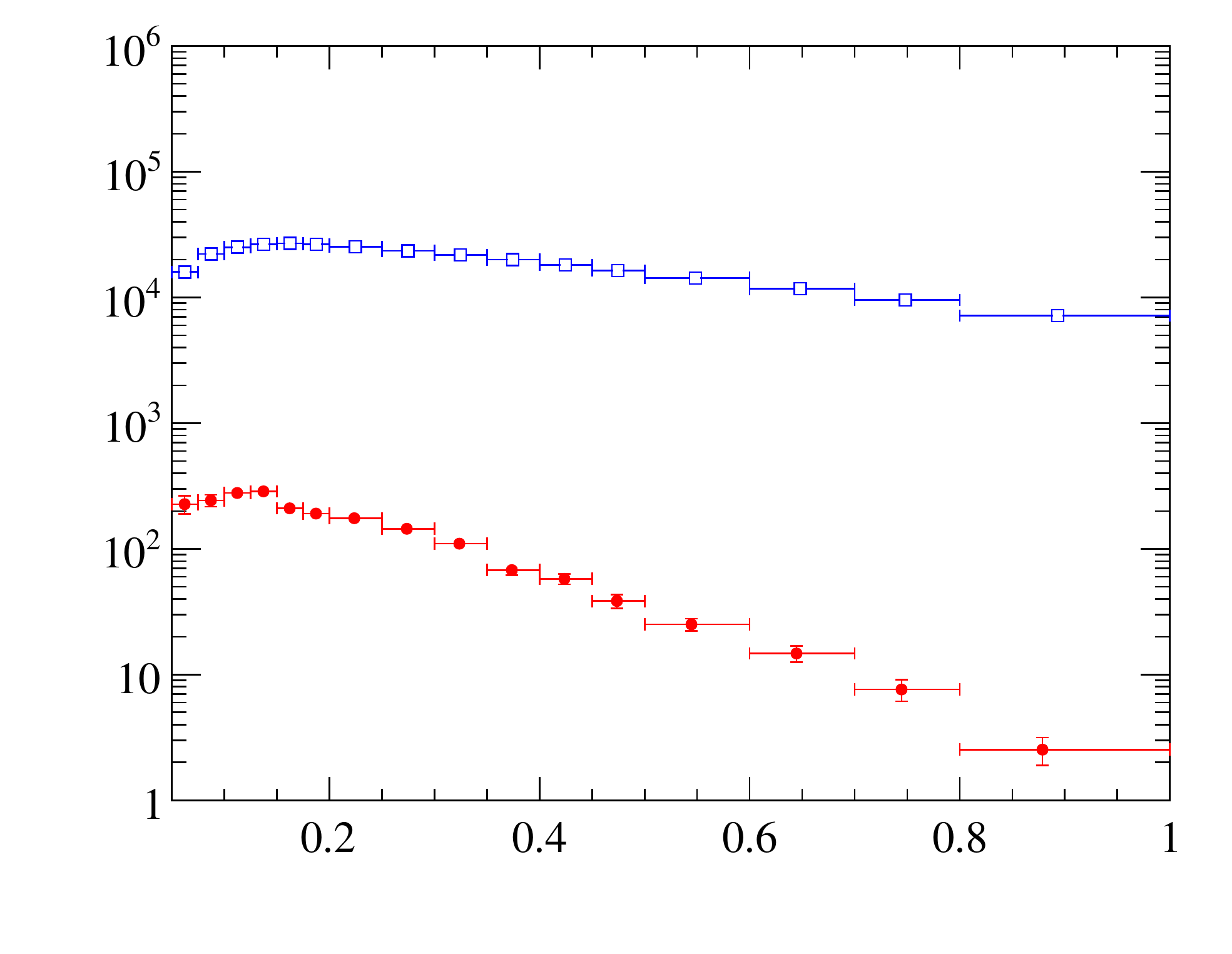}
    }
    \put( 75,  0){
      \includegraphics*[width=75mm,height=60mm,%
      ]{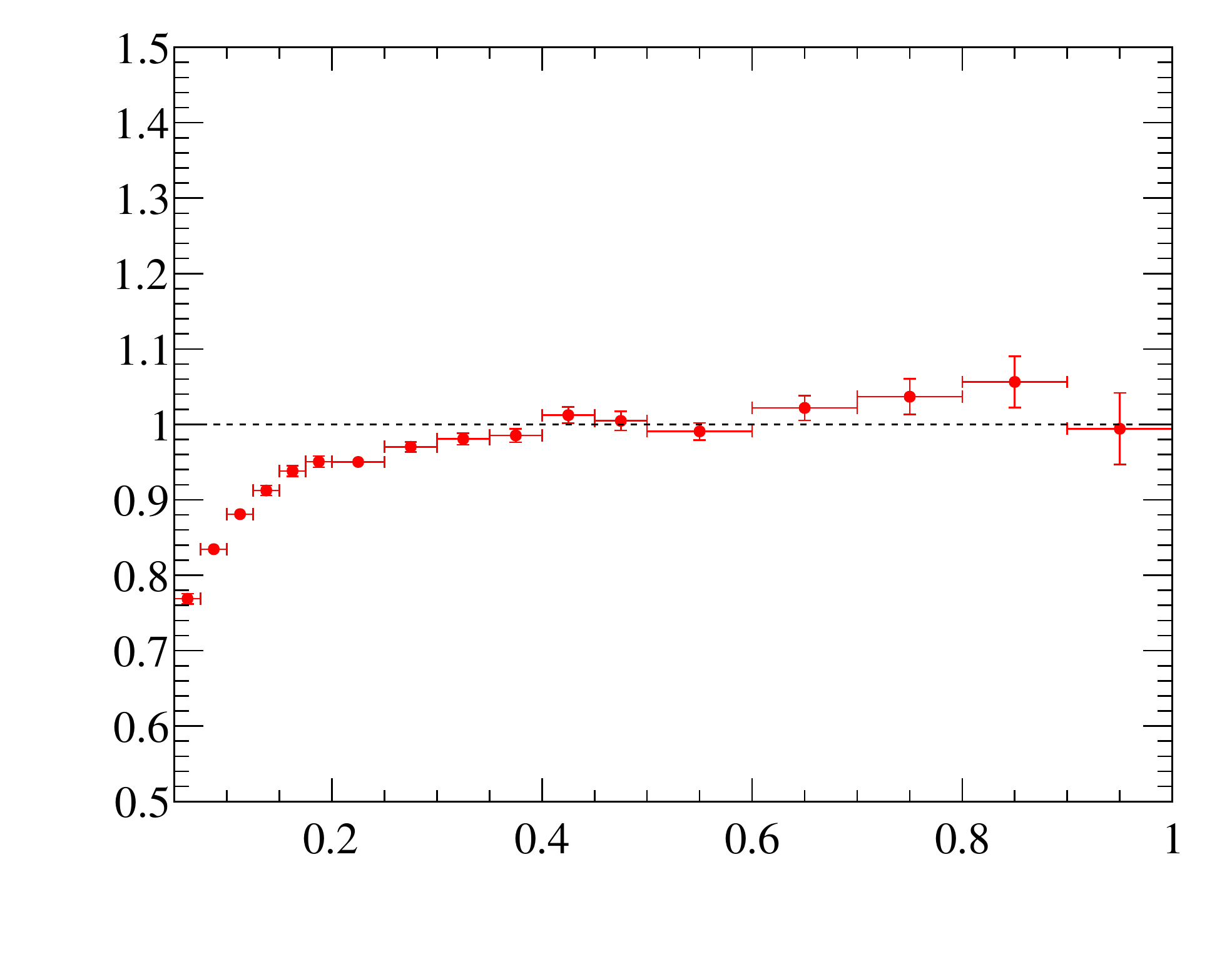}
    }
     \put ( 40, 30) { \small
        $\begin{array}{l}
        {\color{blue}       \square}~~\Bu\to\jpsi\Kp  \\
        {\color{red} \text{\ding{108}}}~~\Bc\to\jpsi\pip        
    \end{array}$}                                      
    \put( 20, 50) {a)}
    \put( 55, 50) {LHCb}
    \put(  0, 30) {\begin{sideways}$\frac{\diff N_{\B}}{\diff t}~~~~\left[\frac{1}{25\mum/c}\right]$\end{sideways} }
    \put( 40,  0) { $t$ }
    \put( 59,  0) { $\left[\!\mm/{\mathit{c}}\right]$ }
    \put( 95, 50) {b)}
    \put(110, 50) {LHCb simulation}
    \put( 75, 46){ \begin{sideways}$\mathcal{R}_{\varepsilon}(t)$\end{sideways} }
    \put(115,  0) { $t$ }
    \put(134,  0) { $\left[\!\mm/{\mathit{c}}\right]$ }
  \end{picture}
  \caption { \small
    (a)\,Decay time distributions for selected 
    \mbox{$\Bc\to\jpsi\pip$}\,(red solid circles) and
    \mbox{$\Bu\to\jpsi\Kp$}\,(blue open squares) decays, with 
    the~data points positioned within the~$t$~bins  according
    to Eq.~(6)  in Ref.~\cite{Lafferty:1994cj};
    (b)\,ratio of acceptance functions $\mathcal{R}_{\varepsilon}(ct)$.
    The~uncertainties are due to sample size only. 
    For visualization purposes the~efficiency ratio is normalized 
    as \mbox{$\mathcal{R}_{\varepsilon}(0.5\mm/{\mathit{c}})=1$}.
  }
  \label{fig:comb}
\end{figure}

The signal yields of \Bc and \Bu~mesons in bins of decay time are shown in Fig.~\ref{fig:comb}(a).
A non-uniform binning scheme is chosen with a~minimal bin width
of $25\mum/c$ at low $t$~increasing to $200\mum/c$ at the~largest decay times,
to keep the \Bc~signal yield above 20 for all $t$~bins.
In the~mass fits of the~individual decay time bins
the~peak positions and mass resolutions are fixed
to the~values 
obtained from the~fit in the~entire region, 
\mbox{$50<t < 1000\mum/c$}.

The  decay time~resolution function is estimated using simulated samples 
and found to be well described by
triple Gaussian functions
with overall rms widths of $10.9\mum/{\mathit{c}}$ and
$11.5\mum/{\mathit{c}}$ for~\Bc and \Bu~decays, 
respectively. The ratio of acceptance functions, $\mathcal{R}_{\varepsilon}(t)$,
is determined using the~simulation and shown in Fig.~\ref{fig:comb}(b).
The~variation in the~acceptance  ratio is caused by the~requirement on
the~\jpsi~decay length imposed in the~trigger and the~subsequent selection. 
The~acceptance is calculated as the~ratio of decay time distributions 
of the~reconstructed and selected simulated events to 
the~theoretical\,(exponential) distributions convolved
with the~resolution function.
This~effectively includes
the~corrections due to resolution effects, 
neglected in Eq.~\eqref{eq:rd}.
It is estimated that any~residual bias
is smaller than 0.1\,\%
in the~range $50 < t <1000\mum/c$.

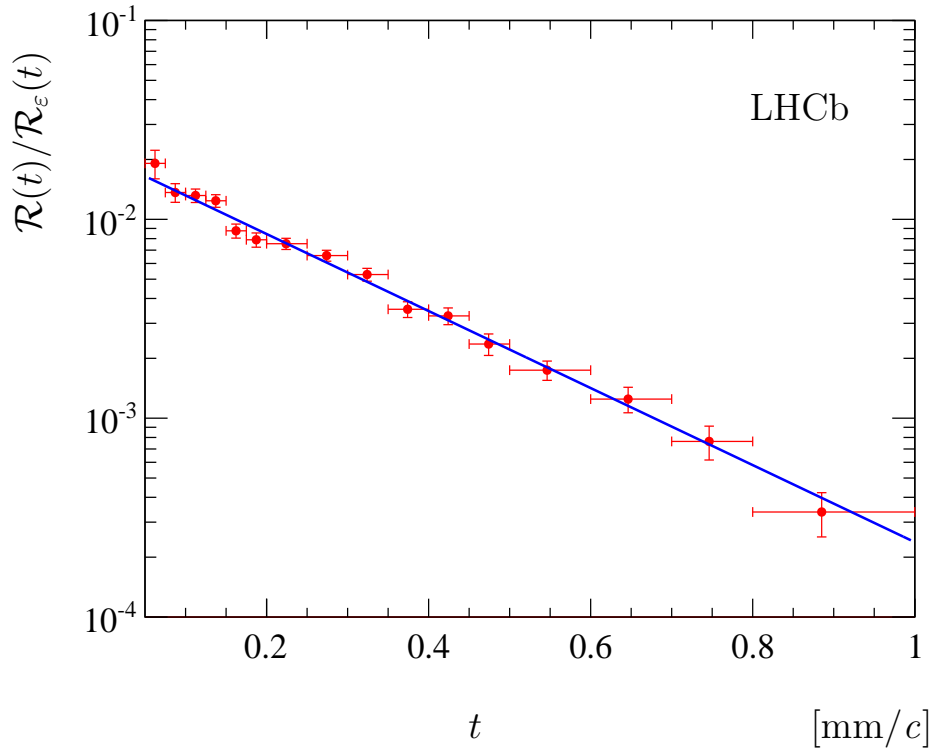
\begin{figure}[t!]
  \setlength{\unitlength}{1mm}
  \centering
  \begin{picture}(125,100)
    \put( 0,0){
      \includegraphics*[width=125mm,height=100mm,%
      ]{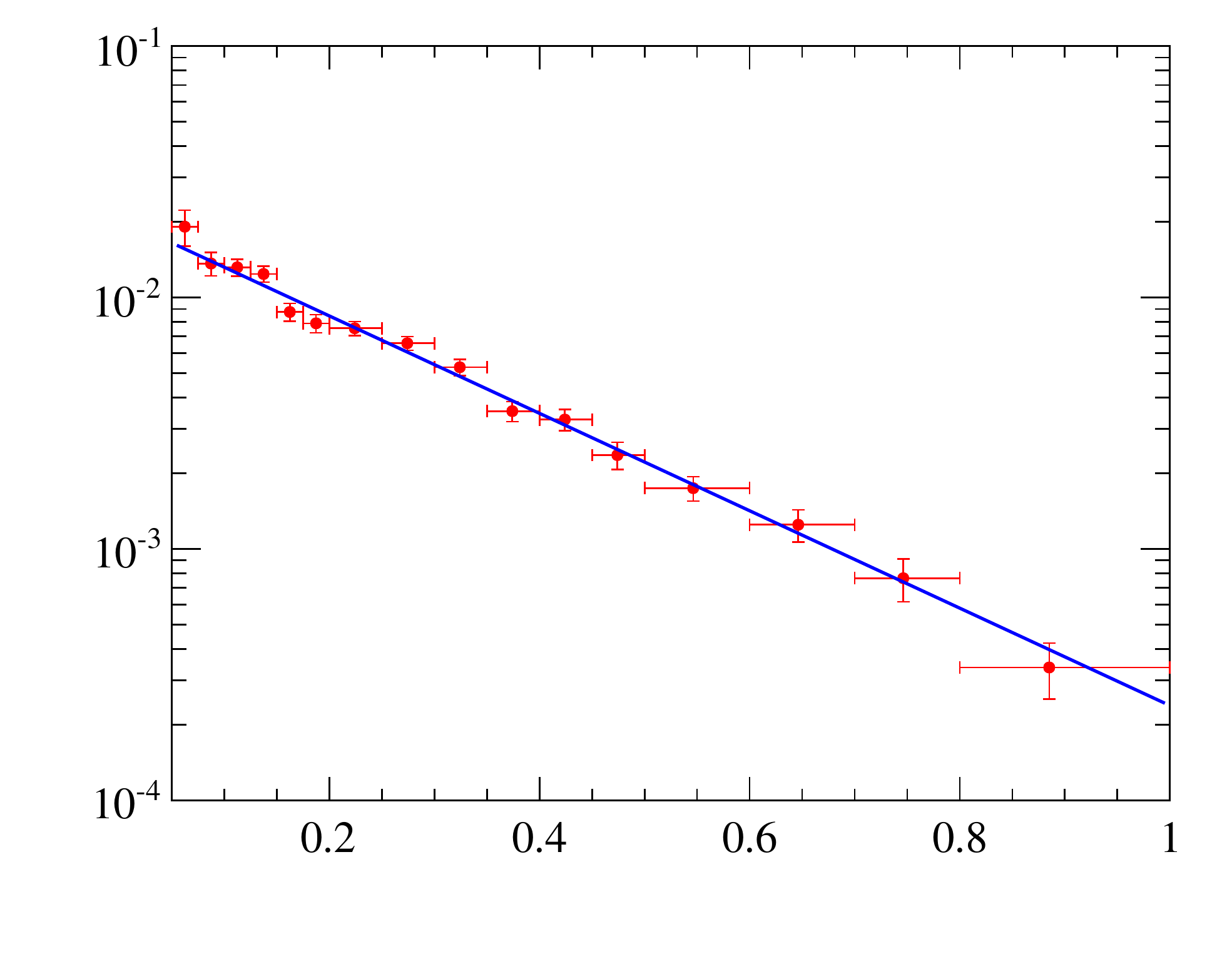}
     }
    \put( 97,82) { \large LHCb}
    \put(  0,67) { \large \begin{sideways}$\mathcal{R}(t)/\mathcal{R}_{\varepsilon}(t)$\end{sideways} }
    \put( 60, 0) { \large $t$ }
    \put(105, 0) { \large $\left[\!\mm/{\mathit{c}}\right]$ }
  \end{picture}
  \caption { \small
    Ratio of the efficiency-corrected decay time distributions\,(points with error bars).
    The~curve shows the~result of the fit with an~exponential function.
    The~data points are positioned within the~$t$~bins  according to Eq.~(6)  in Ref.~\cite{Lafferty:1994cj}.
  }
  \label{fig:deltagamma}
\end{figure}

The efficiency-corrected ratio 
$\mathcal{R}(t)/\mathcal{R}_{\varepsilon}(t)$
is shown in Fig.~\ref{fig:deltagamma}.
A~minimum $\chisq$~fit
with an~exponential function, according to Eq.~\eqref{eq:rd}, gives 
\begin{equation}
    \Delta\Gamma = 4.46 \pm 0.14  \mm^{-1}{\mathit{c}}, \label{eq:dg}
\end{equation}
where the uncertainty is statistical.
The~quality of the~fit is good, with a~$p$-value of~42\,\%.

\vspace*{5mm}

\section{Systematic uncertainties and cross-checks}
\label{sec:syst}
Several sources of systematic uncertainty are considered,
as summarized in Table~\ref{tab:syst} and discussed below. 
\begin{table}[t!]
  \centering
  \caption{
    Summary of systematic uncertainties for $\Delta\Gamma$.
  } \label{tab:syst}
  \vspace*{3mm}
  \begin{tabular*}{0.65\textwidth}{@{\hspace{5mm}}l@{\extracolsep{\fill}}c@{\hspace{5mm}}}
    ~~Source   
    & $\upsigma_{\Delta\Gamma}$ $\left[\mm^{-1}{\mathit{c}}\right]$  
    \\
    \hline 
    Fit model\,(signal and background) &  $0.012$  \\
    $ct$~fit range                     &  $0.040$  \\
    $ct$~binning                       &  $0.016$  \\
    Acceptance                         &  \\
    ~~~Simulation sample size          &  $0.011$  \\ 
    ~~~$\mathtt{MLP}$~filtering        &  $0.025$  \\
    ~~~\jpsi~displacement              &  $0.050$  \\
    \hline  
    Total                              &  $0.072$  
  \end{tabular*}   
\end{table}

The uncertainty related to 
the~determination of the~signal 
yields in $t$~bins is estimated  by comparing 
the nominal results with those obtained using different fit models.
As~an~alternative model
for the~\Bc and \Bu~signals,
a~modified Novosibirsk 
function~\cite{Lees:2011gw} and a~Gaussian function 
are used.
Although the~latter provides poor
description for the~large \Bu~sample for
all decay time bins 
and the~low-background \Bc~signal for 
bins with $t>150\mum/{\mathit{c}}$,
there is no effect
on the~determination of $\Delta\Gamma$.
For the~combinatorial background two alternative parameterizations
are used: 
a~pure exponential function
and a~product of an~exponential function 
and a~second-order polynomial function.
As an~additional check, feed-down components
from the~Cabibbo-suppressed decays 
\mbox{$\Bc\to\jpsi\Kp$}~and
\mbox{$\Bu\to\jpsi\pip$} are added to the~fit.
The~shapes for these components are determined 
using the simulation,
while the~yields are allowed to float 
in the fit. Based on these studies a~systematic uncertainty of 
$0.012\mm^{-1}{\mathit{c}}$ is assigned.
Allowing the position and resolution for 
\Bc and \Bu~signals to vary in fits to
the~individual  $t$~bins does not affect 
the value of $\Delta\Gamma$,
and no systematic uncertainty is  assigned. 

The~uncertainties due to the~choice of $t$~range 
and the~binning scheme are assessed by varying
these and comparing all variants that 
give a~statistical uncertainty for 
$\Delta\Gamma$ below $0.200\mm^{-1}{\mathit{c}}$.
Uncertainties of 
$0.040\mm^{-1}{\mathit{c}}$ and 
$0.016\mm^{-1}{\mathit{c}}$ are assigned due to
the~choice 
of $t$~range and binning scheme. 

The efficiency ratio $\mathcal{R}_{\varepsilon}\left(t\right)$
is determined using simulation, following techniques established in
Refs.~\cite{LHCb-PAPER-2012-017,
  LHCb-PAPER-2013-032,
  LHCb-PAPER-2014-003,
  LHCb-PAPER-2014-021,
  LHCb-PAPER-2014-037,
  LHCb-PAPER-2014-048}.
The uncertainty for $\mathcal{R}_{\varepsilon}\left(t\right)$ due to 
the limited size of the~simulated sample is
estimated to be $0.011\mm^{-1}{\mathit{c}}$
using a~simplified simulation.

The result is stable with respect to large variations 
of the~selection criteria, in particular
the~working points of the~$\mathtt{MLP}$ classifier
and the~displacement criterion for the~\jpsi~vertex.
The~latter is the~only criterion explicitly 
affecting the~lifetime acceptance.
The selection criteria are  varied, 
allowing up to a~20\% increase
in the~statistical
uncertainty for~$\Delta\Gamma$.
Variation of the working point of the~$\mathtt{MLP}$~classifier
results in 
changes of $0.025\mm^{-1}{\mathit{c}}$ in~$\Delta\Gamma$.
Tightening the \jpsi~meson vertex displacement criterion leads to
a~$0.050\mm^{-1}{\mathit{c}}$ change in~$\Delta\Gamma$.
These changes are assigned as systematic uncertainties.
The result is also stable with respect 
to the~choice of the~input variables used
in the~$\mathtt{MLP}$~classifier.
An~alternative selection using
a~boosted decision tree~\cite{Breiman} 
is used for comparison with the $\mathtt{MLP}$ classifier. 
The variation of $\Delta\Gamma$ 
does not exceed a small fraction of its 
statistical uncertainty, 
and no additional systematic uncertainty 
is assigned.
The uncertainties due to the~momentum scale
and the~knowledge of the longitudinal coordinate
of the LHCb vertex detector are studied
in Ref.~\cite{LHCb-PAPER-2013-065}
and found to be negligible.
The~total systematic uncertainty for $\Delta\Gamma$ is 
obtained from the~sum in quadrature of
the~individual contributions
listed in Table~\ref{tab:syst}.

As a~final cross-check, the~whole analysis is repeated using a~lifetime-unbiased
selection, designed to reduce the~lifetime dependence of the~acceptance. 
In this selection, instead of  the~displacement requirements
for the~\jpsi~meson vertex, both at trigger  and subsequent selection,
a~different approach is adopted
requiring the~transverse momentum
of the~\jpsi~meson to be above 3\gevc. 
All other selection criteria are the same,  
including the $\mathtt{MLP}$~classifier. 
This selection has almost uniform acceptance as a~function 
of decay time, but a smaller overall efficiency. 
The~value of $\Delta\Gamma$ obtained using this selection is 
$4.23\pm0.20\mm^{-1}{\mathit{c}}$,
where the~uncertainty is statistical only.
The~larger statistical uncertainty for this 
selection is due to the smaller signal yield and significantly larger background level 
for small $ct$.
The~result agrees with the~baseline selection. 

The results are supported using a~pseudoexperiment technique 
that combines simulation and data.
Each pseudoexperiment is constructed 
from the~sample of \Bu~candidates\,(signal and background) 
from the~data, \ie\ it is the~same for all pseudoexperiments;
the~sample of signal \Bc~mesons is obtained using the~simulation, 
and the~background sample for \Bc~candidates is generated 
using a~simplified simulation according to the 
measured background distributions. 
The~sizes of sub-samples are chosen 
to reproduce the~sample sizes
and background-to-signal ratios for data. 
For each pseudoexperiment 
the~mean lifetime of the~\Bc~meson
is chosen randomly in the~range between 0.6~times and
1.5~times the~known 
\Bc~meson lifetime~\cite{PDG2014}.
The whole analysis is performed 
for each pseudoexperiment and
the~value of 
$\Delta\Gamma$ is determined using 
the~same $\mathcal{R}_{\varepsilon}(t)$~function 
as for the~baseline analysis.
In total 1400 pseudoexperiments are used. 
The~value of $\Delta\Gamma$ 
is found to be unbiased for the~entire
test interval of \Bc~meson lifetimes,
and the error estimate is reliable.

\section{Results and summary}
\label{sec:results}
Using 3.0\invfb of data, collected by the~LHCb experiment 
in $\mathrm{pp}$~collisions at 7~and 8\tev centre-of-mass energies, 
the~difference in total widths between \Bc~and \Bu~mesons is measured to be 
\begin{equation*}
  \Delta\Gamma \equiv
  \Gamma_{\Bc} - \Gamma_{\Bu} =  
    4.46 \pm0.14 \pm0.07 \mm^{-1}{\mathit{c}}, 
\end{equation*}
where the first uncertainty is statistical 
and the~second is systematic.
Using the known lifetime of the \Bu~meson, 
$\uptau_{\Bu}=1.638\pm0.004\ps$~\cite{PDG2014}, 
this is converted 
into a~precise measurement of the~\Bc~meson lifetime,
\begin{equation*}
   \uptau_{\Bc}    =    513.4 \pm 11.0            \pm 5.7\fs,   
\end{equation*}
where in each case the~first uncertainty is statistical, and 
the~second is systematic and includes
the~uncertainty 
related to the~knowledge of the~\Bu~meson lifetime.
This result is in good agreement with the~previous LHCb measurement,
$\uptau_{\Bc}   =   509 \pm 8 \pm 12\fs$,
obtained using semileptonic 
$\Bc\to\jpsi\mup\Pnu_{\upmu}\mathrm{X}$~decays~\cite{LHCb-PAPER-2013-063}, 
and has comparable precision. 
The~uncertainties for these two LHCb measurements 
are uncorrelated, leading to a~combined measurement,
\begin{equation*}
   \uptau_{\Bc}    =    511.4 \pm 9.3\fs,
\end{equation*}
where the statistical and systematic 
uncertainties are added in quadrature.

\section*{Acknowledgements}

\noindent
We express our gratitude to our colleagues in the~CERN
accelerator departments for the excellent performance of the~LHC. We
thank the~technical and administrative staff at the~LHCb
institutes. We acknowledge support from CERN and from the~national
agencies: CAPES, CNPq, FAPERJ and FINEP (Brazil); NSFC (China);
CNRS/IN2P3 (France); BMBF, DFG, HGF and MPG (Germany); INFN (Italy); 
FOM and NWO (The~Netherlands); MNiSW and NCN (Poland); MEN/IFA (Romania); 
MinES and FANO (Russia); MinECo (Spain); SNSF and SER (Switzerland); 
NASU (Ukraine); STFC (United Kingdom); NSF (USA).
The~Tier1 computing centres are supported by IN2P3 (France), KIT and BMBF 
(Germany), INFN (Italy), NWO and SURF (The~Netherlands), PIC (Spain), GridPP 
(United Kingdom).
We are indebted to the~communities behind the~multiple open 
source software packages on which we depend. We are also thankful for
the~computing resources and
the~access to software R\&D tools provided by Yandex LLC (Russia).
Individual groups or members have received support from 
EPLANET, Marie Sk\l{}odowska-Curie Actions and ERC (European Union), 
Conseil g\'{e}n\'{e}ral de Haute-Savoie, Labex ENIGMASS and OCEVU, 
R\'{e}gion Auvergne (France), RFBR (Russia), XuntaGal and GENCAT (Spain), Royal Society and Royal
Commission for
the~Exhibition of 1851 (United Kingdom).
We thank A.K.~Likhoded and A.V.~Luchinsky for fruitful
discussions on \Bc~meson physics.

\addcontentsline{toc}{section}{References}
\setboolean{inbibliography}{true}
\bibliographystyle{LHCb}
\bibliography{main,LHCb-PAPER,LHCb-CONF,LHCb-DP,LHCb-TDR,local}

\ifx\mcitethebibliography\mciteundefinedmacro
\PackageError{LHCb.bst}{mciteplus.sty has not been loaded}
{This bibstyle requires the use of the mciteplus package.}\fi
\providecommand{\href}[2]{#2}
\begin{mcitethebibliography}{10}
\mciteSetBstSublistMode{n}
\mciteSetBstMaxWidthForm{subitem}{\alph{mcitesubitemcount})}
\mciteSetBstSublistLabelBeginEnd{\mcitemaxwidthsubitemform\space}
{\relax}{\relax}

\bibitem{Khoze:1983yp}
V.~A. Khoze and M.~A. Shifman,
  \ifthenelse{\boolean{articletitles}}{\emph{{Heavy quarks}},
  }{}\href{http://dx.doi.org/10.1070/PU1983v026n05ABEH004398}{Sov.\ Phys.\
  Usp.\  \textbf{26} (1983) 387}\relax
\mciteBstWouldAddEndPuncttrue
\mciteSetBstMidEndSepPunct{\mcitedefaultmidpunct}
{\mcitedefaultendpunct}{\mcitedefaultseppunct}\relax
\EndOfBibitem
\bibitem{Bigi:1991ir}
I.~I. Bigi and N.~G. Uraltsev,
  \ifthenelse{\boolean{articletitles}}{\emph{{Gluonic enhancements in
  non-spectator beauty decays: An~inclusive mirage though an~exclusive
  possibility}},
  }{}\href{http://dx.doi.org/10.1016/0370-2693(92)90066-D}{Phys.\ Lett.\
  \textbf{B280} (1992) 271}\relax
\mciteBstWouldAddEndPuncttrue
\mciteSetBstMidEndSepPunct{\mcitedefaultmidpunct}
{\mcitedefaultendpunct}{\mcitedefaultseppunct}\relax
\EndOfBibitem
\bibitem{Blok:1992hw}
B.~Blok and M.~A. Shifman, \ifthenelse{\boolean{articletitles}}{\emph{{The rule
  of discarding $1/N_c$ in inclusive weak decays. 1.}},
  }{}\href{http://dx.doi.org/10.1016/0550-3213(93)90504-I}{Nucl.\ Phys.\
  \textbf{B399} (1993) 441}, \href{http://arxiv.org/abs/hep-ph/9207236}{{\tt
  arXiv:hep-ph/9207236}}\relax
\mciteBstWouldAddEndPuncttrue
\mciteSetBstMidEndSepPunct{\mcitedefaultmidpunct}
{\mcitedefaultendpunct}{\mcitedefaultseppunct}\relax
\EndOfBibitem
\bibitem{Blok:1992he}
B.~Blok and M.~A. Shifman, \ifthenelse{\boolean{articletitles}}{\emph{{The rule
  of discarding $1/N_c$ in inclusive weak decays. 2.}},
  }{}\href{http://dx.doi.org/10.1016/0550-3213(93)90505-J}{Nucl.\ Phys.\
  \textbf{B399} (1993) 459}, \href{http://arxiv.org/abs/hep-ph/9209289}{{\tt
  arXiv:hep-ph/9209289}}\relax
\mciteBstWouldAddEndPuncttrue
\mciteSetBstMidEndSepPunct{\mcitedefaultmidpunct}
{\mcitedefaultendpunct}{\mcitedefaultseppunct}\relax
\EndOfBibitem
\bibitem{Lenz:2014jha}
A.~Lenz, \ifthenelse{\boolean{articletitles}}{\emph{{Lifetimes and HQE}},
  }{}\href{http://arxiv.org/abs/1405.3601}{{\tt arXiv:1405.3601}}\relax
\mciteBstWouldAddEndPuncttrue
\mciteSetBstMidEndSepPunct{\mcitedefaultmidpunct}
{\mcitedefaultendpunct}{\mcitedefaultseppunct}\relax
\EndOfBibitem
\bibitem{PDG2014}
Particle Data Group, K.~A. Olive {\em et~al.},
  \ifthenelse{\boolean{articletitles}}{\emph{{\href{http://pdg.lbl.gov/}{Review
  of particle physics}}},
  }{}\href{http://dx.doi.org/10.1088/1674-1137/38/9/090001}{Chin.\ Phys.\
  \textbf{C38} (2014) 090001}\relax
\mciteBstWouldAddEndPuncttrue
\mciteSetBstMidEndSepPunct{\mcitedefaultmidpunct}
{\mcitedefaultendpunct}{\mcitedefaultseppunct}\relax
\EndOfBibitem
\bibitem{LHCb-PAPER-2013-065}
LHCb collaboration, R.~Aaij {\em et~al.},
  \ifthenelse{\boolean{articletitles}}{\emph{{Measurements of the $B^+$, $B^0$,
  $B_s^0$ meson and $\Lambda_b^0$ baryon lifetimes}},
  }{}\href{http://dx.doi.org/10.1007/JHEP04(2014)114}{JHEP \textbf{04} (2014)
  114}, \href{http://arxiv.org/abs/1402.2554}{{\tt arXiv:1402.2554}}\relax
\mciteBstWouldAddEndPuncttrue
\mciteSetBstMidEndSepPunct{\mcitedefaultmidpunct}
{\mcitedefaultendpunct}{\mcitedefaultseppunct}\relax
\EndOfBibitem
\bibitem{LHCb-PAPER-2013-032}
LHCb collaboration, R.~Aaij {\em et~al.},
  \ifthenelse{\boolean{articletitles}}{\emph{{ measurement of the $\Lambda_b^0$
  baryon lifetime}},
  }{}\href{http://dx.doi.org/10.1103/PhysRevLett.111.102003}{Phys.\ Rev.\
  Lett.\  \textbf{111} (2013) 102003},
  \href{http://arxiv.org/abs/1307.2476}{{\tt arXiv:1307.2476}}\relax
\mciteBstWouldAddEndPuncttrue
\mciteSetBstMidEndSepPunct{\mcitedefaultmidpunct}
{\mcitedefaultendpunct}{\mcitedefaultseppunct}\relax
\EndOfBibitem
\bibitem{LHCb-PAPER-2014-003}
LHCb collaboration, R.~Aaij {\em et~al.},
  \ifthenelse{\boolean{articletitles}}{\emph{{Precision measurement of the
  ratio of the $\Lambda_b^0$ to $\overline{B}^0$ lifetimes}},
  }{}\href{http://dx.doi.org/10.1016/j.physletb.2014.05.021}{Phys.\ Lett.\
  \textbf{B734} (2014) 122}, \href{http://arxiv.org/abs/1402.6242}{{\tt
  arXiv:1402.6242}}\relax
\mciteBstWouldAddEndPuncttrue
\mciteSetBstMidEndSepPunct{\mcitedefaultmidpunct}
{\mcitedefaultendpunct}{\mcitedefaultseppunct}\relax
\EndOfBibitem
\bibitem{LHCb-PAPER-2014-021}
LHCb collaboration, R.~Aaij {\em et~al.},
  \ifthenelse{\boolean{articletitles}}{\emph{{Precision measurement of the mass
  and lifetime of the $\Xi_b^0$ baryon}},
  }{}\href{http://dx.doi.org/10.1103/PhysRevLett.113.032001}{Phys.\ Rev.\
  Lett.\  \textbf{113} (2014) 032001},
  \href{http://arxiv.org/abs/1405.7223}{{\tt arXiv:1405.7223}}\relax
\mciteBstWouldAddEndPuncttrue
\mciteSetBstMidEndSepPunct{\mcitedefaultmidpunct}
{\mcitedefaultendpunct}{\mcitedefaultseppunct}\relax
\EndOfBibitem
\bibitem{LHCb-PAPER-2014-037}
LHCb collaboration, R.~Aaij {\em et~al.},
  \ifthenelse{\boolean{articletitles}}{\emph{{Measurement of the
  $\overline{B}_s^0$ meson lifetime in $D_s^+\pi^-$ decays}},
  }{}\href{http://dx.doi.org/10.1103/PhysRevLett.113.172001}{Phys.\ Rev.\
  Lett.\  \textbf{113} (2014) 172001},
  \href{http://arxiv.org/abs/1407.5873}{{\tt arXiv:1407.5873}}\relax
\mciteBstWouldAddEndPuncttrue
\mciteSetBstMidEndSepPunct{\mcitedefaultmidpunct}
{\mcitedefaultendpunct}{\mcitedefaultseppunct}\relax
\EndOfBibitem
\bibitem{LHCb-PAPER-2014-048}
LHCb collaboration, R.~Aaij {\em et~al.},
  \ifthenelse{\boolean{articletitles}}{\emph{{Precision measurement of the mass
  and lifetime of the $\Xi_b^-$ baryon}},
  }{}\href{http://arxiv.org/abs/1409.8568}{{\tt arXiv:1409.8568}}, {submitted
  to Phys. Rev. Lett.}\relax
\mciteBstWouldAddEndPunctfalse
\mciteSetBstMidEndSepPunct{\mcitedefaultmidpunct}
{}{\mcitedefaultseppunct}\relax
\EndOfBibitem
\bibitem{Chang:1992pt}
C.-H. Chang and Y.-Q. Chen, \ifthenelse{\boolean{articletitles}}{\emph{{The
  decays of $B_{c}$~meson}},
  }{}\href{http://dx.doi.org/10.1103/PhysRevD.49.3399}{Phys.\ Rev.\
  \textbf{D49} (1994) 3399}\relax
\mciteBstWouldAddEndPuncttrue
\mciteSetBstMidEndSepPunct{\mcitedefaultmidpunct}
{\mcitedefaultendpunct}{\mcitedefaultseppunct}\relax
\EndOfBibitem
\bibitem{Gershtein:1994jw}
S.~S. Gershtein, V.~V. Kiselev, A.~K. Likhoded, and A.~V. Tkabladze,
  \ifthenelse{\boolean{articletitles}}{\emph{{Physics of $B_c^+$~mesons}},
  }{}\href{http://dx.doi.org/10.1070/PU1995v038n01ABEH000063}{Phys.\ Usp.\
  \textbf{38} (1995) 1}, \href{http://arxiv.org/abs/hep-ph/9504319}{{\tt
  arXiv:hep-ph/9504319}}\relax
\mciteBstWouldAddEndPuncttrue
\mciteSetBstMidEndSepPunct{\mcitedefaultmidpunct}
{\mcitedefaultendpunct}{\mcitedefaultseppunct}\relax
\EndOfBibitem
\bibitem{Colangelo:1999zn}
P.~Colangelo and F.~De~Fazio, \ifthenelse{\boolean{articletitles}}{\emph{{Using
  heavy quark spin symmetry in semileptonic $B_{c}$~decays}},
  }{}\href{http://dx.doi.org/10.1103/PhysRevD.61.034012}{Phys.\ Rev.\
  \textbf{D61} (2000) 034012}, \href{http://arxiv.org/abs/hep-ph/9909423}{{\tt
  arXiv:hep-ph/9909423}}\relax
\mciteBstWouldAddEndPuncttrue
\mciteSetBstMidEndSepPunct{\mcitedefaultmidpunct}
{\mcitedefaultendpunct}{\mcitedefaultseppunct}\relax
\EndOfBibitem
\bibitem{Kiselev:2000pp}
V.~V. Kiselev, A.~E. Kovalsky, and A.~K. Likhoded,
  \ifthenelse{\boolean{articletitles}}{\emph{{$B_{c}^+$~decays and lifetime in
  QCD sum rules}},
  }{}\href{http://dx.doi.org/10.1016/S0550-3213(00)00386-2}{Nucl.\ Phys.\
  \textbf{B585} (2000) 353}, \href{http://arxiv.org/abs/hep-ph/0002127}{{\tt
  arXiv:hep-ph/0002127}}\relax
\mciteBstWouldAddEndPuncttrue
\mciteSetBstMidEndSepPunct{\mcitedefaultmidpunct}
{\mcitedefaultendpunct}{\mcitedefaultseppunct}\relax
\EndOfBibitem
\bibitem{Kiselev:2001ej}
V.~V. Kiselev, A.~E. Kovalsky, and A.~K. Likhoded,
  \ifthenelse{\boolean{articletitles}}{\emph{{$B_{c}^+$~meson decays and
  lifetime within QCD sum rules}},
  }{}\href{http://dx.doi.org/10.1134/1.1414935}{Phys.\ Atom.\ Nucl.\
  \textbf{64} (2001) 1860}\relax
\mciteBstWouldAddEndPuncttrue
\mciteSetBstMidEndSepPunct{\mcitedefaultmidpunct}
{\mcitedefaultendpunct}{\mcitedefaultseppunct}\relax
\EndOfBibitem
\bibitem{Chang:2000ac}
C.-H. Chang, S.-L. Chen, T.-F. Feng, and X.-Q. Li,
  \ifthenelse{\boolean{articletitles}}{\emph{{The lifetime of $B_{c}^+$~meson
  and some relevant problems}},
  }{}\href{http://dx.doi.org/10.1103/PhysRevD.64.014003}{Phys.\ Rev.\
  \textbf{D64} (2001) 014003}, \href{http://arxiv.org/abs/hep-ph/0007162}{{\tt
  arXiv:hep-ph/0007162}}\relax
\mciteBstWouldAddEndPuncttrue
\mciteSetBstMidEndSepPunct{\mcitedefaultmidpunct}
{\mcitedefaultendpunct}{\mcitedefaultseppunct}\relax
\EndOfBibitem
\bibitem{Chang:2001jn}
C.-H. Chang, S.-L. Chen, T.-F. Fe, and X.-Q. Li,
  \ifthenelse{\boolean{articletitles}}{\emph{{Study of the $B_c$~meson
  lifetime}}, }{}Commun.\ Theor.\ Phys.\  \textbf{35} (2001) 51,
  \href{http://arxiv.org/abs/hep-ph/0103194}{{\tt arXiv:hep-ph/0103194}}\relax
\mciteBstWouldAddEndPuncttrue
\mciteSetBstMidEndSepPunct{\mcitedefaultmidpunct}
{\mcitedefaultendpunct}{\mcitedefaultseppunct}\relax
\EndOfBibitem
\bibitem{Kiselev:2003mp}
V.~V. Kiselev, \ifthenelse{\boolean{articletitles}}{\emph{{Decays of the
  $B_{c}$~meson}}, }{}\href{http://arxiv.org/abs/hep-ph/0308214}{{\tt
  arXiv:hep-ph/0308214}}\relax
\mciteBstWouldAddEndPuncttrue
\mciteSetBstMidEndSepPunct{\mcitedefaultmidpunct}
{\mcitedefaultendpunct}{\mcitedefaultseppunct}\relax
\EndOfBibitem
\bibitem{Kiselev:2001fw}
V.~V. Kiselev and A.~K. Likhoded,
  \ifthenelse{\boolean{articletitles}}{\emph{{Baryons with two heavy quarks}},
  }{}\href{http://dx.doi.org/10.1070/PU2002v045n05ABEH000958}{Phys.\ Usp.\
  \textbf{45} (2002) 455}, \href{http://arxiv.org/abs/hep-ph/0103169}{{\tt
  arXiv:hep-ph/0103169}}\relax
\mciteBstWouldAddEndPuncttrue
\mciteSetBstMidEndSepPunct{\mcitedefaultmidpunct}
{\mcitedefaultendpunct}{\mcitedefaultseppunct}\relax
\EndOfBibitem
\bibitem{Karliner:2014gca}
M.~Karliner and J.~L. Rosner,
  \ifthenelse{\boolean{articletitles}}{\emph{{Baryons with two heavy quarks:
  Masses, production, decays, and detection}},
  }{}\href{http://arxiv.org/abs/1408.5877}{{\tt arXiv:1408.5877}}\relax
\mciteBstWouldAddEndPuncttrue
\mciteSetBstMidEndSepPunct{\mcitedefaultmidpunct}
{\mcitedefaultendpunct}{\mcitedefaultseppunct}\relax
\EndOfBibitem
\bibitem{Abe:1998wi}
CDF collaboration, F.~Abe {\em et~al.},
  \ifthenelse{\boolean{articletitles}}{\emph{{Observation of the $B_{c}$ meson
  in $p\bar{p}$~collisions at $\sqrt{s} = 1.8\tev$}},
  }{}\href{http://dx.doi.org/10.1103/PhysRevLett.81.2432}{Phys.\ Rev.\ Lett.\
  \textbf{81} (1998) 2432}, \href{http://arxiv.org/abs/hep-ex/9805034}{{\tt
  arXiv:hep-ex/9805034}}\relax
\mciteBstWouldAddEndPuncttrue
\mciteSetBstMidEndSepPunct{\mcitedefaultmidpunct}
{\mcitedefaultendpunct}{\mcitedefaultseppunct}\relax
\EndOfBibitem
\bibitem{Abulencia:2006zu}
CDF collaboration, A.~Abulencia {\em et~al.},
  \ifthenelse{\boolean{articletitles}}{\emph{{Measurement of the~$B_{c}^+$
  meson lifetime using $B_{c}^+ \to J/\psi e^{+}\nu$}},
  }{}\href{http://dx.doi.org/10.1103/PhysRevLett.97.012002}{Phys.\ Rev.\ Lett.\
   \textbf{97} (2006) 012002}, \href{http://arxiv.org/abs/hep-ex/0603027}{{\tt
  arXiv:hep-ex/0603027}}\relax
\mciteBstWouldAddEndPuncttrue
\mciteSetBstMidEndSepPunct{\mcitedefaultmidpunct}
{\mcitedefaultendpunct}{\mcitedefaultseppunct}\relax
\EndOfBibitem
\bibitem{Aaltonen:2012yb}
CDF collaboration, T.~Aaltonen {\em et~al.},
  \ifthenelse{\boolean{articletitles}}{\emph{{Measurement of
  the~$B_{c}^{-}$~meson lifetime in the~decay $B_{c}^{-} \rightarrow
  J/\psi\pi^{-}$}},
  }{}\href{http://dx.doi.org/10.1103/PhysRevD.87.011101}{Phys.\ Rev.\
  \textbf{D87} (2013) 011101}, \href{http://arxiv.org/abs/1210.2366}{{\tt
  arXiv:1210.2366}}\relax
\mciteBstWouldAddEndPuncttrue
\mciteSetBstMidEndSepPunct{\mcitedefaultmidpunct}
{\mcitedefaultendpunct}{\mcitedefaultseppunct}\relax
\EndOfBibitem
\bibitem{Abazov:2008rba}
D0 collaboration, V.~M. Abazov {\em et~al.},
  \ifthenelse{\boolean{articletitles}}{\emph{{Measurement of the~lifetime of
  the $B_{c}^{\pm}$ meson in the~semileptonic decay channel}},
  }{}\href{http://dx.doi.org/10.1103/PhysRevLett.102.092001}{Phys.\ Rev.\
  Lett.\  \textbf{102} (2009) 092001},
  \href{http://arxiv.org/abs/0805.2614}{{\tt arXiv:0805.2614}}\relax
\mciteBstWouldAddEndPuncttrue
\mciteSetBstMidEndSepPunct{\mcitedefaultmidpunct}
{\mcitedefaultendpunct}{\mcitedefaultseppunct}\relax
\EndOfBibitem
\bibitem{PDG2012}
Particle Data Group, J.~Beringer {\em et~al.},
  \ifthenelse{\boolean{articletitles}}{\emph{{\href{http://pdg.lbl.gov/}{Review
  of particle physics}}},
  }{}\href{http://dx.doi.org/10.1103/PhysRevD.86.010001}{Phys.\ Rev.\
  \textbf{D86} (2012) 010001}, {and 2013 partial update for the 2014
  edition}\relax
\mciteBstWouldAddEndPuncttrue
\mciteSetBstMidEndSepPunct{\mcitedefaultmidpunct}
{\mcitedefaultendpunct}{\mcitedefaultseppunct}\relax
\EndOfBibitem
\bibitem{LHCb-PAPER-2013-063}
LHCb collaboration, R.~Aaij {\em et~al.},
  \ifthenelse{\boolean{articletitles}}{\emph{{Measurement of the $B_c^+$ meson
  lifetime using $B_c^+\to J/\psi\mu^+\nu_{\mu} X$ decays}},
  }{}\href{http://dx.doi.org/10.1140/epjc/s10052-014-2839-x}{Eur.\ Phys.\ J.\
  \textbf{C74} (2014) 2839}, \href{http://arxiv.org/abs/1401.6932}{{\tt
  arXiv:1401.6932}}\relax
\mciteBstWouldAddEndPuncttrue
\mciteSetBstMidEndSepPunct{\mcitedefaultmidpunct}
{\mcitedefaultendpunct}{\mcitedefaultseppunct}\relax
\EndOfBibitem
\bibitem{LHCb-PAPER-2012-017}
LHCb collaboration, R.~Aaij {\em et~al.},
  \ifthenelse{\boolean{articletitles}}{\emph{{Measurement of the
  $\overline{B}^0_s$ effective lifetime in the $J/\psi f_0(980)$ final state}},
  }{}\href{http://dx.doi.org/10.1103/PhysRevLett.109.152002}{Phys.\ Rev.\
  Lett.\  \textbf{109} (2012) 152002},
  \href{http://arxiv.org/abs/1207.0878}{{\tt arXiv:1207.0878}}\relax
\mciteBstWouldAddEndPuncttrue
\mciteSetBstMidEndSepPunct{\mcitedefaultmidpunct}
{\mcitedefaultendpunct}{\mcitedefaultseppunct}\relax
\EndOfBibitem
\bibitem{Alves:2008zz}
LHCb collaboration, A.~A. Alves~Jr.\ {\em et~al.},
  \ifthenelse{\boolean{articletitles}}{\emph{{The \lhcb detector at the LHC}},
  }{}\href{http://dx.doi.org/10.1088/1748-0221/3/08/S08005}{JINST \textbf{3}
  (2008) S08005}\relax
\mciteBstWouldAddEndPuncttrue
\mciteSetBstMidEndSepPunct{\mcitedefaultmidpunct}
{\mcitedefaultendpunct}{\mcitedefaultseppunct}\relax
\EndOfBibitem
\bibitem{LHCb-DP-2014-001}
R.~Aaij {\em et~al.}, \ifthenelse{\boolean{articletitles}}{\emph{{Performance
  of the LHCb Vertex Locator}},
  }{}\href{http://dx.doi.org/10.1088/1748-0221/9/09/P09007}{JINST \textbf{9}
  (2014) P09007}, \href{http://arxiv.org/abs/1405.7808}{{\tt
  arXiv:1405.7808}}\relax
\mciteBstWouldAddEndPuncttrue
\mciteSetBstMidEndSepPunct{\mcitedefaultmidpunct}
{\mcitedefaultendpunct}{\mcitedefaultseppunct}\relax
\EndOfBibitem
\bibitem{LHCb-DP-2013-003}
R.~Arink {\em et~al.}, \ifthenelse{\boolean{articletitles}}{\emph{{Performance
  of the LHCb Outer Tracker}},
  }{}\href{http://dx.doi.org/10.1088/1748-0221/9/01/P01002}{JINST \textbf{9}
  (2014) P01002}, \href{http://arxiv.org/abs/1311.3893}{{\tt
  arXiv:1311.3893}}\relax
\mciteBstWouldAddEndPuncttrue
\mciteSetBstMidEndSepPunct{\mcitedefaultmidpunct}
{\mcitedefaultendpunct}{\mcitedefaultseppunct}\relax
\EndOfBibitem
\bibitem{LHCb-DP-2012-003}
M.~Adinolfi {\em et~al.},
  \ifthenelse{\boolean{articletitles}}{\emph{{Performance of the \lhcb RICH
  detector at the LHC}},
  }{}\href{http://dx.doi.org/10.1140/epjc/s10052-013-2431-9}{Eur.\ Phys.\ J.\
  \textbf{C73} (2013) 2431}, \href{http://arxiv.org/abs/1211.6759}{{\tt
  arXiv:1211.6759}}\relax
\mciteBstWouldAddEndPuncttrue
\mciteSetBstMidEndSepPunct{\mcitedefaultmidpunct}
{\mcitedefaultendpunct}{\mcitedefaultseppunct}\relax
\EndOfBibitem
\bibitem{LHCb-DP-2012-002}
A.~A. Alves~Jr.\ {\em et~al.},
  \ifthenelse{\boolean{articletitles}}{\emph{{Performance of the LHCb muon
  system}}, }{}\href{http://dx.doi.org/10.1088/1748-0221/8/02/P02022}{JINST
  \textbf{8} (2013) P02022}, \href{http://arxiv.org/abs/1211.1346}{{\tt
  arXiv:1211.1346}}\relax
\mciteBstWouldAddEndPuncttrue
\mciteSetBstMidEndSepPunct{\mcitedefaultmidpunct}
{\mcitedefaultendpunct}{\mcitedefaultseppunct}\relax
\EndOfBibitem
\bibitem{LHCb-DP-2012-004}
R.~Aaij {\em et~al.}, \ifthenelse{\boolean{articletitles}}{\emph{{The \lhcb
  trigger and its performance in 2011}},
  }{}\href{http://dx.doi.org/10.1088/1748-0221/8/04/P04022}{JINST \textbf{8}
  (2013) P04022}, \href{http://arxiv.org/abs/1211.3055}{{\tt
  arXiv:1211.3055}}\relax
\mciteBstWouldAddEndPuncttrue
\mciteSetBstMidEndSepPunct{\mcitedefaultmidpunct}
{\mcitedefaultendpunct}{\mcitedefaultseppunct}\relax
\EndOfBibitem
\bibitem{Sjostrand:2006za}
T.~Sj\"{o}strand, S.~Mrenna, and P.~Skands,
  \ifthenelse{\boolean{articletitles}}{\emph{{{\mbox{\pythia 6.4}} physics and
  manual}}, }{}\href{http://dx.doi.org/10.1088/1126-6708/2006/05/026}{JHEP
  \textbf{05} (2006) 026}, \href{http://arxiv.org/abs/hep-ph/0603175}{{\tt
  arXiv:hep-ph/0603175}}\relax
\mciteBstWouldAddEndPuncttrue
\mciteSetBstMidEndSepPunct{\mcitedefaultmidpunct}
{\mcitedefaultendpunct}{\mcitedefaultseppunct}\relax
\EndOfBibitem
\bibitem{Sjostrand:2007gs}
T.~Sj\"{o}strand, S.~Mrenna, and P.~Skands,
  \ifthenelse{\boolean{articletitles}}{\emph{{A brief introduction to
  {\mbox{\pythia 8.1}}}},
  }{}\href{http://dx.doi.org/10.1016/j.cpc.2008.01.036}{Comput.\ Phys.\
  Commun.\  \textbf{178} (2008) 852},
  \href{http://arxiv.org/abs/0710.3820}{{\tt arXiv:0710.3820}}\relax
\mciteBstWouldAddEndPuncttrue
\mciteSetBstMidEndSepPunct{\mcitedefaultmidpunct}
{\mcitedefaultendpunct}{\mcitedefaultseppunct}\relax
\EndOfBibitem
\bibitem{LHCb-PROC-2010-056}
I.~Belyaev {\em et~al.}, \ifthenelse{\boolean{articletitles}}{\emph{{Handling
  of the generation of primary events in~\gauss, the LHCb simulation
  framework}}, }{}\href{http://dx.doi.org/10.1109/NSSMIC.2010.5873949}{Nuclear
  Science Symposium Conference Record (NSS/MIC) \textbf{IEEE} (2010)
  1155}\relax
\mciteBstWouldAddEndPuncttrue
\mciteSetBstMidEndSepPunct{\mcitedefaultmidpunct}
{\mcitedefaultendpunct}{\mcitedefaultseppunct}\relax
\EndOfBibitem
\bibitem{Chang:2003cq}
C.-H. Chang, C.~Driouichi, P.~Eerola, and X.~G. Wu,
  \ifthenelse{\boolean{articletitles}}{\emph{{\bcvegpy: An event generator for
  hadronic production of the $B_c^+$~meson}},
  }{}\href{http://dx.doi.org/10.1016/j.cpc.2004.02.005}{Comput.\ Phys.\
  Commun.\  \textbf{159} (2004) 192},
  \href{http://arxiv.org/abs/hep-ph/0309120}{{\tt arXiv:hep-ph/0309120}}\relax
\mciteBstWouldAddEndPuncttrue
\mciteSetBstMidEndSepPunct{\mcitedefaultmidpunct}
{\mcitedefaultendpunct}{\mcitedefaultseppunct}\relax
\EndOfBibitem
\bibitem{Chang:2005hq}
C.-H. Chang, J.-X. Wang, and X.-G. Wu,
  \ifthenelse{\boolean{articletitles}}{\emph{{\bcvegpy\,2.0: An~upgrade version
  of the generator \bcvegpy with an addendum about hadroproduction of the
  P-wave $B_c^+$~states}},
  }{}\href{http://dx.doi.org/10.1016/j.cpc.2005.09.008}{Comput.\ Phys.\
  Commun.\  \textbf{174} (2006) 241},
  \href{http://arxiv.org/abs/hep-ph/0504017}{{\tt arXiv:hep-ph/0504017}}\relax
\mciteBstWouldAddEndPuncttrue
\mciteSetBstMidEndSepPunct{\mcitedefaultmidpunct}
{\mcitedefaultendpunct}{\mcitedefaultseppunct}\relax
\EndOfBibitem
\bibitem{Wu:2013pya}
X.-G. Wu, \ifthenelse{\boolean{articletitles}}{\emph{{\bcvegpy and
  {\sc{GenXicc}} for the hadronic production of the doubly heavy mesons and
  baryons}}, }{}\href{http://dx.doi.org/10.1088/1742-6596/523/1/012042}{J.\
  Phys.\ Conf.\ Ser.\  \textbf{523} (2014) 012042},
  \href{http://arxiv.org/abs/1307.3344}{{\tt arXiv:1307.3344}}\relax
\mciteBstWouldAddEndPuncttrue
\mciteSetBstMidEndSepPunct{\mcitedefaultmidpunct}
{\mcitedefaultendpunct}{\mcitedefaultseppunct}\relax
\EndOfBibitem
\bibitem{Berezhnoy:1994ba}
A.~V. Berezhnoy, A.~K. Likhoded, and M.~V. Shevlyagin,
  \ifthenelse{\boolean{articletitles}}{\emph{{Hadronic production of
  $B_c^+$~mesons}}, }{}Phys.\ Atom.\ Nucl.\  \textbf{58} (1995) 672,
  \href{http://arxiv.org/abs/hep-ph/9408284}{{\tt arXiv:hep-ph/9408284}}\relax
\mciteBstWouldAddEndPuncttrue
\mciteSetBstMidEndSepPunct{\mcitedefaultmidpunct}
{\mcitedefaultendpunct}{\mcitedefaultseppunct}\relax
\EndOfBibitem
\bibitem{Kolodziej:1995nv}
{K.\ ~Kolodziej, A.\ ~Leike and R.\ ~R\"uckl},
  \ifthenelse{\boolean{articletitles}}{\emph{{Production of $B_c^{+}$~mesons in
  hadronic collisions}},
  }{}\href{http://dx.doi.org/10.1016/0370-2693(95)00710-3}{Phys.\ Lett.\
  \textbf{B355} (1995) 337}, \href{http://arxiv.org/abs/hep-ph/9505298}{{\tt
  arXiv:hep-ph/9505298}}\relax
\mciteBstWouldAddEndPuncttrue
\mciteSetBstMidEndSepPunct{\mcitedefaultmidpunct}
{\mcitedefaultendpunct}{\mcitedefaultseppunct}\relax
\EndOfBibitem
\bibitem{Chang:1994aw}
C.-H. Chang, Y.-Q. Chen, G.-P. Han, and H.-T. Jiang,
  \ifthenelse{\boolean{articletitles}}{\emph{{On hadronic production of the
  $B_c^+$~meson}},
  }{}\href{http://dx.doi.org/10.1016/0370-2693(95)01235-4}{Phys.\ Lett.\
  \textbf{B364} (1995) 78}, \href{http://arxiv.org/abs/hep-ph/9408242}{{\tt
  arXiv:hep-ph/9408242}}\relax
\mciteBstWouldAddEndPuncttrue
\mciteSetBstMidEndSepPunct{\mcitedefaultmidpunct}
{\mcitedefaultendpunct}{\mcitedefaultseppunct}\relax
\EndOfBibitem
\bibitem{LHCb-PAPER-2013-010}
LHCb collaboration, R.~Aaij {\em et~al.},
  \ifthenelse{\boolean{articletitles}}{\emph{{Observation of $B^+_c \to J/\psi
  D_s^+$ and $B^+_c \to J/\psi D_s^{*+}$ decays}},
  }{}\href{http://dx.doi.org/10.1103/PhysRevD.87.112012}{Phys.\ Rev.\
  \textbf{D87} (2013) 112012}, \href{http://arxiv.org/abs/1304.4530}{{\tt
  arXiv:1304.4530}}\relax
\mciteBstWouldAddEndPuncttrue
\mciteSetBstMidEndSepPunct{\mcitedefaultmidpunct}
{\mcitedefaultendpunct}{\mcitedefaultseppunct}\relax
\EndOfBibitem
\bibitem{LHCb-PAPER-2013-021}
LHCb collaboration, R.~Aaij {\em et~al.},
  \ifthenelse{\boolean{articletitles}}{\emph{{First observation of the decay
  $B_c^+ \to J/\psi K^+$}},
  }{}\href{http://dx.doi.org/10.1007/JHEP09(2013)075}{JHEP \textbf{09} (2013)
  075}, \href{http://arxiv.org/abs/1306.6723}{{\tt arXiv:1306.6723}}\relax
\mciteBstWouldAddEndPuncttrue
\mciteSetBstMidEndSepPunct{\mcitedefaultmidpunct}
{\mcitedefaultendpunct}{\mcitedefaultseppunct}\relax
\EndOfBibitem
\bibitem{LHCb-PAPER-2013-044}
LHCb collaboration, R.~Aaij {\em et~al.},
  \ifthenelse{\boolean{articletitles}}{\emph{{Observation of the decay
  $B_c^+\to B_s^0\pi^+$}},
  }{}\href{http://dx.doi.org/10.1103/PhysRevLett.111.181801}{Phys.\ Rev.\
  Lett.\  \textbf{111} (2013) 181801},
  \href{http://arxiv.org/abs/1308.4544}{{\tt arXiv:1308.4544}}\relax
\mciteBstWouldAddEndPuncttrue
\mciteSetBstMidEndSepPunct{\mcitedefaultmidpunct}
{\mcitedefaultendpunct}{\mcitedefaultseppunct}\relax
\EndOfBibitem
\bibitem{LHCb-PAPER-2013-047}
LHCb collaboration, R.~Aaij {\em et~al.},
  \ifthenelse{\boolean{articletitles}}{\emph{{Observation of the decay $B_c^+
  \to J/\psi K^+K^-\pi^+$}},
  }{}\href{http://dx.doi.org/10.1007/JHEP11(2013)094}{JHEP \textbf{11} (2013)
  094}, \href{http://arxiv.org/abs/1309.0587}{{\tt arXiv:1309.0587}}\relax
\mciteBstWouldAddEndPuncttrue
\mciteSetBstMidEndSepPunct{\mcitedefaultmidpunct}
{\mcitedefaultendpunct}{\mcitedefaultseppunct}\relax
\EndOfBibitem
\bibitem{LHCb-PAPER-2014-009}
LHCb collaboration, R.~Aaij {\em et~al.},
  \ifthenelse{\boolean{articletitles}}{\emph{{Evidence for the decay $B_c^+ \to
  J/\psi 3\pi^+ 2\pi^-$}},
  }{}\href{http://dx.doi.org/10.1007/JHEP05(2014)148}{JHEP \textbf{05} (2014)
  148}, \href{http://arxiv.org/abs/1404.0287}{{\tt arXiv:1404.0287}}\relax
\mciteBstWouldAddEndPuncttrue
\mciteSetBstMidEndSepPunct{\mcitedefaultmidpunct}
{\mcitedefaultendpunct}{\mcitedefaultseppunct}\relax
\EndOfBibitem
\bibitem{LHCb-PAPER-2014-025}
LHCb collaboration, R.~Aaij {\em et~al.},
  \ifthenelse{\boolean{articletitles}}{\emph{{Measurement of the ratio of
  $B_c^+$ branching fractions to $J/\psi\pi^+$ and $J/\psi\mu^+\nu_\mu$}},
  }{}\href{http://dx.doi.org/10.1103/PhysRevD.90.032009}{Phys.\ Rev.\
  \textbf{D90} (2014) 032009}, \href{http://arxiv.org/abs/1407.2126}{{\tt
  arXiv:1407.2126}}\relax
\mciteBstWouldAddEndPuncttrue
\mciteSetBstMidEndSepPunct{\mcitedefaultmidpunct}
{\mcitedefaultendpunct}{\mcitedefaultseppunct}\relax
\EndOfBibitem
\bibitem{LHCb-PAPER-2014-039}
LHCb collaboration, R.~Aaij {\em et~al.},
  \ifthenelse{\boolean{articletitles}}{\emph{{First observation of a baryonic
  $B_c^+$ decay}},
  }{}\href{http://dx.doi.org/10.1103/PhysRevLett.113.152003}{Phys.\ Rev.\
  Lett.\  \textbf{113} (2014) 152003},
  \href{http://arxiv.org/abs/1408.0971}{{\tt arXiv:1408.0971}}\relax
\mciteBstWouldAddEndPuncttrue
\mciteSetBstMidEndSepPunct{\mcitedefaultmidpunct}
{\mcitedefaultendpunct}{\mcitedefaultseppunct}\relax
\EndOfBibitem
\bibitem{LHCb-PAPER-2014-050}
LHCb collaboration, R.~Aaij {\em et~al.},
  \ifthenelse{\boolean{articletitles}}{\emph{{Measurement of $B_c^+$ production
  at $\sqrt{s}=8$ TeV}}, }{}\href{http://arxiv.org/abs/1411.2943}{{\tt
  arXiv:1411.2943}}, {submitted to Phys. Rev. Lett.}\relax
\mciteBstWouldAddEndPunctfalse
\mciteSetBstMidEndSepPunct{\mcitedefaultmidpunct}
{}{\mcitedefaultseppunct}\relax
\EndOfBibitem
\bibitem{Lange:2001uf}
D.~J. Lange, \ifthenelse{\boolean{articletitles}}{\emph{{The \evtgen particle
  decay simulation package}},
  }{}\href{http://dx.doi.org/10.1016/S0168-9002(01)00089-4}{Nucl.\ Instrum.\
  Meth.\  \textbf{A462} (2001) 152}\relax
\mciteBstWouldAddEndPuncttrue
\mciteSetBstMidEndSepPunct{\mcitedefaultmidpunct}
{\mcitedefaultendpunct}{\mcitedefaultseppunct}\relax
\EndOfBibitem
\bibitem{Golonka:2005pn}
P.~Golonka and Z.~Was, \ifthenelse{\boolean{articletitles}}{\emph{{\photos
  Monte Carlo: A precision tool for QED corrections in $Z$ and $W$ decays}},
  }{}\href{http://dx.doi.org/10.1140/epjc/s2005-02396-4}{Eur.\ Phys.\ J.\
  \textbf{C45} (2006) 97}, \href{http://arxiv.org/abs/hep-ph/0506026}{{\tt
  arXiv:hep-ph/0506026}}\relax
\mciteBstWouldAddEndPuncttrue
\mciteSetBstMidEndSepPunct{\mcitedefaultmidpunct}
{\mcitedefaultendpunct}{\mcitedefaultseppunct}\relax
\EndOfBibitem
\bibitem{Allison:2006ve}
Geant4 collaboration, J.~Allison {\em et~al.},
  \ifthenelse{\boolean{articletitles}}{\emph{{\geant developments and
  applications}}, }{}\href{http://dx.doi.org/10.1109/TNS.2006.869826}{IEEE
  Trans.\ Nucl.\ Sci.\  \textbf{53} (2006) 270}\relax
\mciteBstWouldAddEndPuncttrue
\mciteSetBstMidEndSepPunct{\mcitedefaultmidpunct}
{\mcitedefaultendpunct}{\mcitedefaultseppunct}\relax
\EndOfBibitem
\bibitem{Agostinelli:2002hh}
Geant4 collaboration, S.~Agostinelli {\em et~al.},
  \ifthenelse{\boolean{articletitles}}{\emph{{\geant: A~simulation toolkit}},
  }{}\href{http://dx.doi.org/10.1016/S0168-9002(03)01368-8}{Nucl.\ Instrum.\
  Meth.\  \textbf{A506} (2003) 250}\relax
\mciteBstWouldAddEndPuncttrue
\mciteSetBstMidEndSepPunct{\mcitedefaultmidpunct}
{\mcitedefaultendpunct}{\mcitedefaultseppunct}\relax
\EndOfBibitem
\bibitem{LHCb-PROC-2011-006}
M.~Clemencic {\em et~al.}, \ifthenelse{\boolean{articletitles}}{\emph{{The
  \lhcb simulation application, \gauss: Design, evolution and experience}},
  }{}\href{http://dx.doi.org/10.1088/1742-6596/331/3/032023}{{J.\ Phys.\ Conf.\
  Ser.\ } \textbf{331} (2011) 032023}\relax
\mciteBstWouldAddEndPuncttrue
\mciteSetBstMidEndSepPunct{\mcitedefaultmidpunct}
{\mcitedefaultendpunct}{\mcitedefaultseppunct}\relax
\EndOfBibitem
\bibitem{McCulloch}
W.~S. McCulloch and W.~Pitts, \ifthenelse{\boolean{articletitles}}{\emph{A
  logical calculus of the ideas immanent in nervous activity},
  }{}\href{http://dx.doi.org/10.1007/BF02478259}{The bulletin of mathematical
  biophysics \textbf{5} (1943), no.~4 115}\relax
\mciteBstWouldAddEndPuncttrue
\mciteSetBstMidEndSepPunct{\mcitedefaultmidpunct}
{\mcitedefaultendpunct}{\mcitedefaultseppunct}\relax
\EndOfBibitem
\bibitem{rosenblatt58}
F.~Rosenblatt, \ifthenelse{\boolean{articletitles}}{\emph{The perceptron:
  A~probabilistic model for information storage and organization in the brain},
  }{}Psychological Review \textbf{65} (1958) 386\relax
\mciteBstWouldAddEndPuncttrue
\mciteSetBstMidEndSepPunct{\mcitedefaultmidpunct}
{\mcitedefaultendpunct}{\mcitedefaultseppunct}\relax
\EndOfBibitem
\bibitem{Zhong:2011xm}
J.-H. Zhong {\em et~al.}, \ifthenelse{\boolean{articletitles}}{\emph{{A~program
  for the Bayesian neural network in the \root~framework}},
  }{}\href{http://dx.doi.org/10.1016/j.cpc.2011.07.019}{Comput.\ Phys.\
  Commun.\  \textbf{182} (2011) 2655},
  \href{http://arxiv.org/abs/1103.2854}{{\tt arXiv:1103.2854}}\relax
\mciteBstWouldAddEndPuncttrue
\mciteSetBstMidEndSepPunct{\mcitedefaultmidpunct}
{\mcitedefaultendpunct}{\mcitedefaultseppunct}\relax
\EndOfBibitem
\bibitem{Hulsbergen:2005pu}
W.~D. Hulsbergen, \ifthenelse{\boolean{articletitles}}{\emph{{Decay chain
  fitting with a Kalman filter}},
  }{}\href{http://dx.doi.org/10.1016/j.nima.2005.06.078}{Nucl.\ Instrum.\
  Meth.\  \textbf{A552} (2005) 566},
  \href{http://arxiv.org/abs/physics/0503191}{{\tt
  arXiv:physics/0503191}}\relax
\mciteBstWouldAddEndPuncttrue
\mciteSetBstMidEndSepPunct{\mcitedefaultmidpunct}
{\mcitedefaultendpunct}{\mcitedefaultseppunct}\relax
\EndOfBibitem
\bibitem{LHCb-PAPER-2011-013}
LHCb collaboration, R.~Aaij {\em et~al.},
  \ifthenelse{\boolean{articletitles}}{\emph{{Observation of $J/\psi$-pair
  production in pp collisions at $\sqrt{s}=7$ TeV}},
  }{}\href{http://dx.doi.org/10.1016/j.physletb.2011.12.015}{Phys.\ Lett.\
  \textbf{B707} (2012) 52}, \href{http://arxiv.org/abs/1109.0963}{{\tt
  arXiv:1109.0963}}\relax
\mciteBstWouldAddEndPuncttrue
\mciteSetBstMidEndSepPunct{\mcitedefaultmidpunct}
{\mcitedefaultendpunct}{\mcitedefaultseppunct}\relax
\EndOfBibitem
\bibitem{Lafferty:1994cj}
G.~D. Lafferty and T.~R. Wyatt,
  \ifthenelse{\boolean{articletitles}}{\emph{{Where to stick your data points:
  The treatment of measurements within wide bins}},
  }{}\href{http://dx.doi.org/10.1016/0168-9002(94)01112-5}{Nucl.\ Instrum.\
  Meth.\  \textbf{A355} (1995) 541}\relax
\mciteBstWouldAddEndPuncttrue
\mciteSetBstMidEndSepPunct{\mcitedefaultmidpunct}
{\mcitedefaultendpunct}{\mcitedefaultseppunct}\relax
\EndOfBibitem
\bibitem{Lees:2011gw}
BaBar collaboration, J.~P. Lees {\em et~al.},
  \ifthenelse{\boolean{articletitles}}{\emph{{Branching fraction measurements
  of the color-suppressed becays $ \overline{B}^0 \to D^{(*)0} \pi^0$,
  $D^{(*)0} \eta$, $D^{(*)0} \omega$, and $D^{(*)0} \eta^{\prime}$ and
  measurement of the polarization in the decay $\overline{B}^0 \to D^{*0}
  \omega$}}, }{}\href{http://dx.doi.org/10.1103/PhysRevD.84.112007}{Phys.\
  Rev.\  \textbf{D84} (2011) 112007}, Erratum
  \href{http://dx.doi.org/10.1103/PhysRevD.87.039901}{ibid.\   \textbf{D87}
  (2013) 039901}, \href{http://arxiv.org/abs/1107.5751}{{\tt
  arXiv:1107.5751}}\relax
\mciteBstWouldAddEndPuncttrue
\mciteSetBstMidEndSepPunct{\mcitedefaultmidpunct}
{\mcitedefaultendpunct}{\mcitedefaultseppunct}\relax
\EndOfBibitem
\bibitem{Breiman}
L.~Breiman, J.~H. Friedman, R.~A. Olshen, and C.~J. Stone, {\em Classification
  and regression trees}, Wadsworth International Group, Belmont, California,
  USA, 1984\relax
\mciteBstWouldAddEndPuncttrue
\mciteSetBstMidEndSepPunct{\mcitedefaultmidpunct}
{\mcitedefaultendpunct}{\mcitedefaultseppunct}\relax
\EndOfBibitem
\end{mcitethebibliography}

\newpage
\centerline{\large\bf LHCb collaboration}
\begin{flushleft}
\small
R.~Aaij$^{41}$, 
B.~Adeva$^{37}$, 
M.~Adinolfi$^{46}$, 
A.~Affolder$^{52}$, 
Z.~Ajaltouni$^{5}$, 
S.~Akar$^{6}$, 
J.~Albrecht$^{9}$, 
F.~Alessio$^{38}$, 
M.~Alexander$^{51}$, 
S.~Ali$^{41}$
% ...

{\footnotesize \it
% ...
$ ^{5}$Clermont Universit\'{e}, Universit\'{e} Blaise Pascal, CNRS/IN2P3, LPC, Clermont-Ferrand, France\\
$ ^{6}$CPPM, Aix-Marseille Universit\'{e}, CNRS/IN2P3, Marseille, France\\
% ...
$ ^{9}$Fakult\"{a}t Physik, Technische Universit\"{a}t Dortmund, Dortmund, Germany\\
% ...
$ ^{37}$Universidad de Santiago de Compostela, Santiago de Compostela, Spain\\
$ ^{38}$European Organization for Nuclear Research (CERN), Geneva, Switzerland\\
% ...
$ ^{41}$Nikhef National Institute for Subatomic Physics, Amsterdam, The Netherlands\\
% ...
$ ^{46}$H.H. Wills Physics Laboratory, University of Bristol, Bristol, United Kingdom\\
% ...
$ ^{51}$School of Physics and Astronomy, University of Glasgow, Glasgow, United Kingdom\\
$ ^{52}$Oliver Lodge Laboratory, University of Liverpool, Liverpool, United Kingdom\\
% ...
}
\end{flushleft}



\end{document}